\documentclass[12pt,draftcls,onecolumn]{IEEEtran}
\usepackage[utf8]{inputenc}
\usepackage{amsmath,mathtools}
\usepackage{amsthm}
\usepackage{amsfonts}
\usepackage{amssymb}
\usepackage{cite} 
\usepackage{array} 

\usepackage{caption}
\usepackage{url}
\usepackage[list=true]{subcaption}
\usepackage{easyReview}
\usepackage{soulutf8}
\newcolumntype{C}[1]{>{\centering\let\newline\\\arraybackslash\hspace{0pt}}m{#1}}
\DeclarePairedDelimiter\abs{\lvert}{\rvert}

\DeclarePairedDelimiter\curlybracket{\lbrace}{\rbrace}
\newcommand{\Exp}[1]{\ensuremath{\operatorname{e}^{#1}}}
\begin{document}
	\bstctlcite{IEEEexample:BSTcontrol}
	\title{Product and Ratio of Two $\alpha-\kappa - \mu $ Shadowed Random Variables and its Application to Wireless Communication}
	\author{Shashank Shekhar, Sheetal Kalyani \\
		\hspace{-0.5 cm}Department of Electrical Engineering,\\
		\hspace{-0.8 cm} Indian Institute of Technology, Madras, \\
		\hspace{-1cm} Chennai, India 600036.\\
		\hspace{-1cm} \{ee17d022@smail,skalyani@ee\}.iitm.ac.in\\
	}
	\maketitle
		\begin{abstract}
		This work studies the product and ratio statistics of independent and non-identically distributed (i.n.i.d) $ \alpha-\kappa - \mu $ shadowed random variables. We derive the series expression for the probability density function (PDF), cumulative distribution function (CDF), and moment generating function (MGF) of the product and ratio of i.n.i.d $ \alpha - \kappa - \mu $ shadowed random variables. We then give the single integral representation for the derived PDF expressions. Further, as application examples, 1) outage probability has been derived for cascaded wireless systems, and 2) physical-layer security metrics like secrecy outage probability and strictly positive secrecy capacity are derived for the classic three-node model with $\alpha-\kappa-\mu$ shadowed fading. Next, we discuss an intelligent reflecting surface-assisted communication system over $\alpha-\kappa-\mu$ shadowed fading.
	\end{abstract}
	
	
	\section{Introduction}\label{sec:introduction}
	A wireless channel is governed mainly by two physical phenomena: shadowing (which results in long-term signal variation) and multipath (which results in short-term fading). Shadowing is typically modeled using lognormal distribution \cite{alouini2002dual} or sometimes using gamma distribution \cite{abdi1999utility}. In contrast to shadowing, the multipath effect is characterized by a broad range of distributions such as Rayleigh, Rician, Nakagami-$ m $, Hoyt, and more general distributions like $ \kappa-\mu $, $ \eta-\mu $ \cite{Yacoub2007:KappaMu}. The $  \kappa - \mu $ shadowed distribution introduced in \cite{paris2013:statistical} provides a natural generalization of the $ \kappa - \mu $ fading where the line-of-sight (LOS) component of the received signal is random, \textit{i.e.}, subject to shadowing. This type of fading model is known as the LOS shadow fading model in the literature.
	\par All the fading models mentioned above, including $  \kappa - \mu $ shadowed and its special cases, assume a homogeneous scattering environment where the intensity of the received signal is the sum of the intensity of different multipath components \cite{yacoub2007alpha}. To generalize these models and characterize the non-linearity of the propagation medium, a new parameter $\alpha$ was introduced in \cite{yacoub2007alpha}.
	Recently, a generalization of $  \kappa - \mu $ shadowed distribution is discussed in \cite{ramirez2019alpha} that incorporates the non-linearity of the propagation environment. This fading distribution is known as $\alpha-\kappa-\mu$ shadowed distribution where the parameter $\alpha \in \mathbb{R}^{+} $ determines the non-linearity of the propagation environment.
	\par In the literature, Fisher - Snedecor $ \mathcal{F} $  distribution \cite{yoo2017fisher} is also suggested for modeling small-scale fading. The properties of Fisher - Snedecor $ \mathcal{F} $  distribution and its related distributions are discussed in \cite{yoo2017fisher,du2019distribution}. The physical model of  Fisher - Snedecor $ \mathcal{F} $  distribution is similar to the Nakagami-$m$ fading model with modification to incorporate the shadowing effect by multiplying an inverse Nakagami-$m$ random variable (RV). However, this fading model results in a simple form for probability density function (PDF) and cumulative distribution function (CDF) expression. This fading model does not include the various generalized fading models such as $ \kappa - \mu  $ fading, Rician fading, and others as special cases. Also, the Fisher - Snedecor $ \mathcal{F} $  distribution does not incorporate the non-linearity of the propagation environment, which is modeled in $\alpha - \kappa - \mu $ shadowed fading by parameter $\alpha$. 
	\par In a variety of wireless communication applications, such as relay-based communication systems \cite{talha2011channel},  and intelligent reflecting surfaces (IRS) assisted communication system \cite{lis_renzo_2020,wu2019towards,Zhang5g}, the transmitted signal from the source reaches the destination after experiencing a couple of fading environments. To analyze such a communication system's performance, one needs to know the statistics of the product of corresponding fading distributions. Hence, the statistical characterization of the product of two RVs is crucial in wireless communication. Similarly, the ratio of random variables appears in scenarios like the study of signal-to-interference ratio and secrecy analysis. A frequency domain based approach is presented in \cite{peppas2015physical} to evaluate the metrics related to physical layer security in wireless systems. The study of product and ratio of probability distribution has been an area of interest for statisticians for a long time. The product and ratio statistics of  $H$-distribution were studied in \cite{carter1977distribution}.
	For example, in bistatic scatter radio communication, the indirect channel between carrier emitter and software-defined radio (SDR) reader through a radio frequency (RF)  tag is modeled as the product of two Rayleigh and Rician fading channels in \cite{fasarakis2015coherent} and \cite{alevizos2017noncoherent}, respectively.  A close approximation for the product of independent Rayleigh RVs is proposed in \cite{lu2011accurate}. The work in \cite{fasarakis2015coherent} and \cite{alevizos2017noncoherent} were focused on point-to-point communication.  In contrast, a multi-scatter scenario is considered in \cite{alevizos2018multistatic} where multiple carrier emitters are present, and each channel is modeled as Nakagami$-m$ fading. Hence, the channel between the carrier emitter and the SDR  reader is a product of two Nakagami$-m$ RVs. In \cite{salo2006distribution}, authors presented a general result for the product statistics of Rayleigh fading. A generic cascaded channel has been considered in \cite{karagiannidis2007n,yilmaz2009product} with Nakagami$-m$ and generalized Nakagami$-m$ fading, respectively. An IRS-assisted wireless communication system under Nakagami$-m$ fading was studied in \cite{selimis2021performance}.
	The authors in \cite{Charishma2021OutageIRS} considered an IRS-assisted communication system where each link undergoes $ \kappa-\mu $ fading; hence, the link between source to IRS and IRS to destination is the product of two $ \kappa-\mu $ RVs. Authors in \cite{Bhargav2018:ProductKappaMu} studied the product statistics of two independent
	and non-identically distributed (i.n.i.d.) $ \kappa-\mu $ RVs. In \cite{Silva:ProductAlphaKappaEtaMu} the statistical characterization of the product of two  i.n.i.d. $ \alpha-\mu $, $ \eta-\mu $ and $ \kappa-\mu $ RVs is done. The author in \cite{bilim2022cascaded} studied the product statistics of two i.n.i.d. $ \kappa - \mu $ shadowed RVs.
	\par In this work, we are interested in $ \alpha - \kappa - \mu $ shadowed fading since it encompasses various popular fading models such as one-sided Gaussian, Rayleigh, Rician, Nakagami-$m$, $ \kappa - \mu $, $\eta-\mu$ \cite{Yacoub2007:KappaMu}, Rician shadowed \cite{abdi2003new} and $\kappa-\mu$ shadowed \cite{paris2013:statistical}. Other than this, the $ \alpha - \kappa - \mu $ shadowed fading has received a lot of traction in the wireless community in recent literature like \cite{sarker2021intercept,badrudduza2021secrecy,bhardwaj2021fixed}. Recently, authors in \cite{wu2023accurate} approximated the $ \alpha - \kappa - \mu $ shadowed RV with generalized Gamma RV and used it to analyze secure ultra-reliable and low latency communications systems.  This motivates us to look at the product and ratio statistics of $ \alpha-\kappa - \mu $ shadowed fading. We provided the statistical characterization of the product and ratio of two independent non-identically distributed (i.n.i.d.)  $ \alpha-\kappa - \mu $ shadowed RVs. The closed-form exact expressions for PDF and CDF of the considered RVs are derived using Mellin transformation.    
	The contribution and utility of this work are summarized as follows:
	\begin{itemize}
		\item Series expressions for PDF, CDF, and MGF of the product and ratio of two, i.n.i.d. $ \alpha-\kappa-\mu $ shadowed RV are derived using the direct application of Mellin transform.
		\item We have derived a novel Laplace-type integral representation for the derived PDF expression of the product and ratio statistics.
		\item We presented the performance metrics for a cascaded wireless system, physical layer security, and outage probability (OP) for an IRS-assisted wireless communication system as application examples for the derived theoretical results to highlight the utility of this work.
		\item The results derived here generalize many existing results in the literature. For example, the cascade wireless system results presented in \cite{bilim2022cascaded,Bhargav2018:ProductKappaMu,Silva:ProductAlphaKappaEtaMu} can be easily deduced from our results. Similarly, our results on physical layer security generalize works in \cite{belmoubarik2014secrecy,iwata2017secure,bhargav2016secrecy,srinivasan2018secrecy,badrudduza2021secrecy}. The OP expressions derived for an IRS-assisted wireless communication system in \cite{tao2020performance,kudathanthirige_icc_20,atapattu2020reconfigurable} are the special cases of our derived expression.
	\end{itemize}
	\section{Definitions and Problem Statement}\label{Sec:  ProductAKMS_Def}
	In $ \alpha-\kappa-\mu $ shadowed fading environment, the received signal envelope $ R $ is as follows \cite{ramirez2019alpha}
	\begin{equation}
		\begin{aligned}
			R^{\alpha} &= \sum_{i=1}^{\mu}\left[ \left(P_{i} + \xi p_{i}\right)^{2} + \left(Q_{i} + \xi q_{i}\right)^{2}\right],
		\end{aligned}
	\end{equation}
	where $ \alpha \in \mathbb{R}^{+} $ characterizes the non-linearity of the propagation medium. $ \mu$ denotes the number of multipath clusters. In each cluster, the scattered components $ P_{i},Q_{i} $ are modeled as $ \mathcal{N}\left(0,\sigma^{2} \right) $. The dominant components are characterized by real numbers $ p_{i}, q_{i} $, and RV $ \xi $ that follows Nakagami-$m$ distribution with shape parameter $m$ and spread parameter $ \Omega = 1 $. The set of parameters $ \{ \alpha,\kappa,\mu,m  \} $ completely defines the $ \alpha-\kappa-\mu  $ shadowed fading model. The parameter $ \kappa = \sum\limits_{i=1}^{\mu}\left( p_{i}^{2} + q_{i}^{2} \right)/2\sigma^{2}$ signifies the ratio of dominant component power and scattered component power.
	
	We considered two independent non-identically distributed (i.n.i.d) $ \alpha-\kappa-\mu $ shadowed RVs, say $ X_{i} $ with mean $ \bar{\gamma}_{i} $ and non-negative real shaping parameters $\alpha_{i}, \kappa_{i},\mu_{i}, m_{i} $ for $ i = 1,2 $. Each $ X_{i} $ follows the distribution given by\cite[Eq. $(4)$]{ramirez2019alpha},
	\begin{equation}
		\label{Eq: ProductAKMS_AKappaMuSpdf}
		\begin{aligned}
			f_{X_{i}}(x_{i}) &= \frac{ \theta_{i}  \alpha_{i} }{2 c_{i}^{\mu_{i}}  \bar{\gamma}_{i}}\left(\frac{x_{i}}{\bar{\gamma}_{i}}\right)^{\frac{\alpha_{i} \mu_{i}}{2}-1}  \exp\left({-\frac{1}{c_{i}}\left(\frac{x_{i}}{\bar{\gamma}_{i}}\right)^{\frac{\alpha_{i}}{2}} }\right) \\
			&\times { }_{1} F_{1}\left(m_{i}; \mu_{i} ; \frac{\beta_{i}}{c_{i}} \left(\frac{x_{i}}{\bar{\gamma}_{i}}\right)^{\frac{\alpha_{i}}{2}} \right),
		\end{aligned}
	\end{equation}
	where $ i = 1,2 $, $ \theta_{i} = \frac{m_{i}^{m_{i}}}{\Gamma\left(\mu_{i}\right) \left(\mu_{i} \kappa_{i}+m_{i}\right)^{m_{i}} },  \beta_{i} = \frac{ \mu_{i} \kappa_{i}}{\left(\mu_{i} \kappa_{i} + m_{i}\right)}  $, $ {}_{1} F_{1}\left(\cdot;\cdot;\cdot\right) $ is confluent hypergeometric function\cite{srivastava1985:multiple} and 
	\begin{equation}
		\begin{aligned}
			c_{i} &= \left(\frac{1}{\theta_{i} \Gamma\left(\mu_{i} + \frac{2}{\alpha_{i}}\right) {}_{2}F_{1}\left(m_{i},\mu_{i} + \frac{2}{\alpha_{i}} ; \mu_{i};\beta_{i} \right) }\right)^{\frac{\alpha_{i}}{2}},
		\end{aligned}
	\end{equation}
	where, $ {}_{2} F_{1}\left(\cdot,\cdot;\cdot;\cdot\right) $ is Gauss hypergeometric function\cite{srivastava1985:multiple}. Our objective is to statistically characterize the RV  $ Y = X_{1} X_{2} $ and $ Z = \frac{X_{1}}{X_{2}} $. The following sections will provide the analytical expression for PDF, CDF, and MGF of RV $ Y $ and $ Z $.         
	\section{Statistical Characterization of Product Statistics}\label{Sec: ProductAKMS_product}
	In this section, We derive the probability density function (PDF), cumulative distribution function (CDF), and moment generating function (MGF) of RV $ Y $, \textit{i.e.,} the product of two $ \alpha-\kappa-\mu $ shadowed RVs by using the technique of Mellin transform. From the definition of Mellin transform \cite[Eq. (2.8.9)]{springer1979:RValgebra} and \cite[Eq. (8)]{ramirez2019alpha}, we have 
	\begin{equation}
		\begin{aligned}
			\mathcal{M}\left[f_{X_{i}}(x_{i});s\right] &= \theta_{i} \bar{\gamma}_{i}^{s-1} c_{i}^{\frac{2\left(s-1\right)}{\alpha_{i}}}  \Gamma\left(\mu_{i} + \left(s-1\right)\frac{2}{\alpha_{i}}\right) \\&\times {}_{2}F_{1}\left(m_{i},\mu_{i} + \left(s-1\right)\frac{2}{\alpha_{i}};\mu_{i};\beta_{i}\right),
		\end{aligned}
	\end{equation}
	\par It is a well-known fact that the Mellin convolution of individual PDFs gives the PDF of the product of two independent RVs, and the Mellin transform of the said PDF is the product of the Mellin transform of corresponding PDFs \cite{springer1979:RValgebra}. Hence, the Mellin transform of $ Y $ is 
	\begin{equation}\label{Eq: ProductAKMS_MellinTransY}
		\begin{aligned}
			\mathcal{M}\left[f_{Y}\left(y\right);s\right] &= \mathcal{M}\left[f_{X_{1}}(x_{1});s\right]  \mathcal{M}\left[f_{X_{2}}(x_{2});s\right] \\
			&= \theta_{1}\theta_{2} \left(\bar{\gamma}_{1}\bar{\gamma}_{2}\right)^{s-1} c_{1}^{\frac{2\left(s-1\right)}{\alpha_{1}}} c_{2}^{\frac{2\left(s-1\right)}{\alpha_{2}}}   \\
			&\hspace{-4em}\times  \Gamma\left(\mu_{1} + \left(s-1\right)\frac{2}{\alpha_{1}}\right) \Gamma\left(\mu_{2} + \left(s-1\right)\frac{2}{\alpha_{2}}\right)  \\
			&\hspace{-4em}\times{}_{2}F_{1}\left(m_{1},\mu_{1} + \left(s-1\right)\frac{2}{\alpha_{1}};\mu_{1};\beta_{1}\right)  \\
			&\hspace{-4em}\times {}_{2}F_{1}\left(m_{2},\mu_{2} + \left(s-1\right)\frac{2}{\alpha_{2}};\mu_{2};\beta_{2}\right).
		\end{aligned}
	\end{equation}
	Note that $ \beta_{1},\beta_{2} < 1 $ and $ \mu_{1},\mu_{2} > 0 $ so both the hypergeometric function present in the Mellin transform of $ Y $ converges $ \forall s $. Using the series expansion of hypergeometric function \cite[Eq. (1.1.18)]{srivastava1985:multiple}, we have
	\begin{equation}\label{Eq: ProductAKMS_MellinTransYSeries}
		\begin{aligned}
			\mathcal{M}\left[f_{Y}\left(y\right);s\right] &= \frac{\theta_{1}\theta_{2} c_{1}^{-\frac{2}{\alpha_{1}}} c_{2}^{-\frac{2}{\alpha_{2}}} }{\bar{\gamma}_{1}\bar{\gamma}_{2}}  \sum_{u,v=0}^{\infty} \frac{\left(m_{1}\right)_{u} \left(m_{2}\right)_{v} \beta_{1}^{u} \beta_{2}^{v} }{\left(\mu_{1}\right)_{u} \left(\mu_{2}\right)_{v} u!v!} \\&\times  \Gamma\left(\mu_{1} + u - \frac{2}{\alpha_{1}} +  \frac{2}{\alpha_{1}}s\right)\\ &\times   \Gamma\left(\mu_{2} + v - \frac{2}{\alpha_{2}} + \frac{2}{\alpha_{2}}s\right) \left(\frac{c_{1}^{-\frac{2}{\alpha_{1}}} c_{2}^{-\frac{2}{\alpha_{2}}} }{\bar{\gamma}_{1}\bar{\gamma}_{2}   }\right)^{-s}
		\end{aligned}
	\end{equation}  
	Now, $ f_{Y}\left(y\right) $ is obtained using the inverse Mellin transform \cite[Eq. 2.8]{mathai2009h}, \textit{i.e.,}
	\begin{equation}\label{Eq: ProductAKMS_ExactProdPDF}
		\begin{aligned}
			f_{Y}\left(y\right) &= \delta\theta_{1}\theta_{2}  \sum_{u,v=0}^{\infty} A_{u,v} H_{0,2}^{2,0}\left[\delta y \  \begin{array}{|c}
				-\\
				\left(b_{1},\frac{2}{\alpha_{1}}\right), \left(b_{2},\frac{2}{\alpha_{2}}\right)
			\end{array}  \right],
		\end{aligned}
	\end{equation}
	where $ A_{u,v} = \frac{\left(m_{1}\right)_{u} \left(m_{2}\right)_{v} \beta_{1}^{u} \beta_{2}^{v} }{\left(\mu_{1}\right)_{u} \left(\mu_{2}\right)_{v} u!v!},  \delta = \frac{c_{1}^{-\frac{2}{\alpha_{1}}} c_{2}^{-\frac{2}{\alpha_{2}}}  }{\bar{\gamma}_{1}\bar{\gamma}_{2}}, b_{1} = \mu_{1}+u-\frac{2}{\alpha_{1}}, b_{2} = \mu_{2}+v-\frac{2}{\alpha_{2}} $, and $ H_{p,q}^{m,n}\left[x \ \begin{array}{|c}
		\left(a_{p},A_{p}\right) \\
		\left(b_{q},B_{q}\right)
	\end{array}\right] $ is the H-function defined in \cite[Eq. 1.1]{mathai2009h}. Equation (\ref{Eq: ProductAKMS_ExactProdPDF}) provides the PDF of the product of two $ \alpha-\kappa - \mu $ shadowed RV in the form of double infinite summation. 
	\subsection{Cumulative Distribution Function}
	The CDF of RV $ Y $, \textit{i.e.}, $ F_{Y}\left(y\right) =  \int\limits_{0}^{y} f_{Y}\left(t\right) \operatorname{dt} $ follows from the definition of H-function and given as  
	\begin{equation}
		\begin{aligned}
			F_{Y}\left(y\right) &= \delta\theta_{1}\theta_{2}  \sum_{u,v=0}^{\infty} A_{u,v} \frac{1}{2\pi i} \int_{\mathcal{C}} \Gamma\left(b_{1}+ \frac{2}{\alpha_{1}}s \right) \\ & \times   \Gamma\left(b_{2} + \frac{2}{\alpha_{2}}s \right) \left( \int\limits_{0}^{y} \left(\delta t\right)^{-s}  \operatorname{dt} \right) \operatorname{ds}
		\end{aligned}
	\end{equation}
	After solving the inner integral and substituting the limits, we have 
	\begin{equation}\label{Eq: ProductAKMS_ExactProdCDF}
		\begin{aligned}
			F_{Y}\left(y\right) &= \delta\theta_{1}\theta_{2} y \sum_{u,v=0}^{\infty} A_{u,v} \\ & \times   H_{1,3}^{2,1}\left[\delta y \  \begin{array}{|c}
				\left(-1,1\right)\\
				\left(b_{1},\frac{2}{\alpha_{1}}\right), \left(b_{2},\frac{2}{\alpha_{2}}\right), \left(0,1\right)
			\end{array}  \right], \\
			&\hspace{-2em}\stackrel{\left(a\right)}{=} \theta_{1}\theta_{2}  \sum_{u,v=0}^{\infty}A_{u,v} H_{1,3}^{2,1}\left[\delta y \  \begin{array}{|c}
				\left(0,1\right)\\
				\left(b_{3},\frac{2}{\alpha_{1}}\right), \left(b_{4},\frac{2}{\alpha_{2}}\right), \left(1,1\right)
			\end{array}  \right],
		\end{aligned}
	\end{equation}
	where $  b_{3} = \mu_{1}+u, b_{4} = \mu_{2}+v $ and $ \left(a\right) $ follows from a functional relation given in \cite[Property 1.5, Eq. 1.60]{mathai2009h}. Next, we derive the closed-form/analytical expression for the MGF of RV $ Y $.
	\subsection{Moment Generating Function}
	The moment generating function (MGF) of an RV is directly related to the Laplace transform of PDF as
	\begin{equation}\label{Eq: ProductAKMS_MGFDef}
		\begin{aligned}
			M_{Y}\left(s\right) &= \mathcal{L}\left[f_{Y}\left(y\right);-s\right] .
		\end{aligned}
	\end{equation} 
	Substituting $ (\ref{Eq: ProductAKMS_ExactProdCDF}) $ in $ (\ref{Eq: ProductAKMS_MGFDef}) $ and then using the identity \cite[Eq. 2.19]{mathai2009h}, we have 
	\begin{equation}
		\begin{aligned}
			M_{Y}\left(s\right) 
			&= \theta_{1}\theta_{2}  \sum_{u,v=0}^{\infty}A_{u,v} \\& \times H_{1,2}^{2,1}\left[\frac{\delta}{\left(-s\right)} \  \begin{array}{|c}
				\left(1,1\right)\\
				\left(b_{3},\frac{2}{\alpha_{1}}\right), \left(b_{4},\frac{2}{\alpha_{2}}\right)
			\end{array}  \right],
		\end{aligned}
	\end{equation}
	\subsection{Higher Moments of $ Y $}
	The $ n $-th  order moment of RV $ Y $ is given by $ \mathbb{E}\left[Y^{n}\right] = \mathbb{E}\left[X_{1}^{n}\right]\mathbb{E}\left[X_{2}^{n}\right]$, since $ X_{1} $ and $ X_{2} $ are independent RVs. Also, note that the $ \mathbb{E}\left[Y^{n}\right] = \mathcal{M}\left[f_{Y}\left(y\right);s\right]\vert_{s = n+1} $. Hence, from (\ref{Eq: ProductAKMS_MellinTransY}) we have
	\begin{equation}\label{Eq: ProductAKMS_ProductMoments}
		\begin{aligned}
			\mathbb{E}\left[Y^{n}\right] &= 
			\theta_{1}\theta_{2} \left(\bar{\gamma}_{1}\bar{\gamma}_{2}\right)^{n} c_{1}^{\frac{2n}{\alpha_{1}}} c_{2}^{\frac{2n}{\alpha_{2}}}  \Gamma\left(\mu_{1} + \frac{2n}{\alpha_{1}}\right) \\&\times \Gamma\left(\mu_{2} + \frac{2n}{\alpha_{2}}\right)  {}_{2}F_{1}\left(m_{1},\mu_{1} + \frac{2n}{\alpha_{1}};\mu_{1};\beta_{1}\right) \\
			&\times {}_{2}F_{1}\left(m_{2},\mu_{2} + \frac{2n}{\alpha_{2}};\mu_{2};\beta_{2}\right).
		\end{aligned}
	\end{equation}
	\section{Statistical Characterization of Ratio Statistics}\label{Sec: ProductAKMS_ratio}
	In this section, We derive the PDF, CDF, and MGF of RV $ Z $, \textit{i.e.,} the ratio of two $ \alpha-\kappa-\mu $ shadowed RVs. Note that the Mellin transform of the ratio of two independent RVs is related to their individual Mellin transform as follows
	\begin{equation}\label{Eq: ProductKMS_MellinTransRatio} 
		\begin{aligned}
			\mathcal{M}\left[f_{Z}\left(z\right);s\right] &= \mathcal{M}\left[f_{X_{1}}\left(x_{1}\right);s\right] \mathcal{M}\left[f_{X_{2}}\left(x_{2}\right);2-s\right], \\
			&= \theta_{1}\theta_{2} \left(\bar{\gamma}_{1}\right)^{s-1}\left(\bar{\gamma}_{2}\right)^{1-s} c_{1}^{\frac{2\left(s-1\right)}{\alpha_{1}}} c_{2}^{\frac{2\left(1-s\right)}{\alpha_{2}}} \\
			&\hspace{-4em}\times \Gamma\left(\mu_{1} + \left(s-1\right)\frac{2}{\alpha_{1}}\right) \Gamma\left(\mu_{2} + \left(1-s\right)\frac{2}{\alpha_{2}}\right) \\
			&\hspace{-4em}\times {}_{2}F_{1}\left(m_{1},\mu_{1} + \left(s-1\right)\frac{2}{\alpha_{1}};\mu_{1};\beta_{1}\right) \\
			&\hspace{-4em}\times {}_{2}F_{1}\left(m_{2},\mu_{2} + \left(1-s\right)\frac{2}{\alpha_{2}};\mu_{2};\beta_{2}\right).
		\end{aligned}
	\end{equation}
	Again using the series expansion of $ {}_{2}F_{1}\left(\cdot,\cdot;\cdot;\cdot\right) $, we have
	\begin{equation}
		\begin{aligned}
			\mathcal{M}\left[f_{Z}\left(z\right);s\right] &= \frac{\theta_{1}\theta_{2} \bar{\gamma}_{2} c_{1}^{-\frac{2}{\alpha_{1}}} c_{2}^{\frac{2}{\alpha_{2}}} }{\bar{\gamma}_{1}}  \sum_{u,v=0}^{\infty}\frac{\left(m_{1}\right)_{u} \left(m_{2}\right)_{v} \beta_{1}^{u} \beta_{2}^{v} }{\left(\mu_{1}\right)_{u} \left(\mu_{2}\right)_{v} u!v!} \\&\hspace{-4em} \times \Gamma\left(\mu_{1} + u - \frac{2}{\alpha_{1}} +  \frac{2}{\alpha_{1}}s\right)\\ &\hspace{-4em}\times   \Gamma\left(\mu_{2} + v + \frac{2}{\alpha_{2}} - \frac{2}{\alpha_{2}}s\right) \left( \frac{\bar{\gamma}_{2} c_{1}^{-\frac{2}{\alpha_{1}}} c_{2}^{\frac{2}{\alpha_{2}}} }{\bar{\gamma}_{1}  }\right)^{-s}
		\end{aligned}
	\end{equation}
	Now, $ f_{Z}\left(z\right) $ is obtained using the inverse Mellin transform \cite[Eq. 2.8]{mathai2009h}, \textit{i.e.,}
	\begin{equation}\label{Eq: ProductAKMS_ExactRatioPDF}
		\begin{aligned}
			f_{Z}\left(z\right) &= \zeta\theta_{1}\theta_{2}  \sum_{u,v=0}^{\infty}A_{u,v} H_{1,1}^{1,1}\left[\zeta z \  \begin{array}{|c}
				\left(a_{1},\frac{2}{\alpha_{2}}\right)\\[2mm]
				\left(b_{1},\frac{2}{\alpha_{1}}\right)
			\end{array}  \right],
		\end{aligned}
	\end{equation}
	where $ \zeta = \frac{\bar{\gamma}_{2} c_{1}^{-\frac{2}{\alpha_{1}}} c_{2}^{\frac{2}{\alpha_{2}}} }{\bar{\gamma}_{1}  } $ and $ a_{1} = 1 - \mu_{2} - v - \frac{2}{\alpha_{2}} $. Following the similar steps as in the calculation of $F_{Y}\left(y\right)$, we have
	\begin{equation}\label{Eq: ProductAKMS_ExactRatioCDF}
		\begin{aligned}
			F_{Z}\left(z\right) &= \theta_{1}\theta_{2}  \sum_{u,v=0}^{\infty} A_{u,v} H_{2,2}^{1,2}\left[\zeta z \  \begin{array}{|c}
				\left(a_{2},\frac{2}{\alpha_{2}}\right), \left(0,1\right)\\[2mm]
				\left(b_{3},\frac{2}{\alpha_{1}}\right),\left(1,1\right)
			\end{array}  \right],
		\end{aligned}
	\end{equation}
	where $ a_{2} = \mu_{2} + v $, and the MGF of $ Z $ follows from \cite[Eq. 2.19]{mathai2009h} and given as
	\begin{equation}
		\begin{aligned}
			M_{Z}\left(s\right)  &= \theta_{1}\theta_{2}  \sum_{u,v=0}^{\infty} A_{u,v} H_{1,1}^{1,1}\left[\frac{\zeta}{\left(-s\right)} \  \begin{array}{|c}
				\left(1,1\right),\left(a_{2},\frac{2}{\alpha_{2}}\right)\\[2mm]
				\left(b_{3},\frac{2}{\alpha_{1}}\right)
			\end{array}  \right].
		\end{aligned}
	\end{equation}
	\subsection{Higher Moments of $ Z $}
	Unlike the case of product statistics, the $ n $-th order moment of RV $ Z $, \textit{i.e.,} $ \mathbb{E}\left[Z^{n}\right] \ne \frac{\mathbb{E}\left[X_{1}^{n}\right]}{\mathbb{E}\left[X_{2}^{n}\right]} $ even though the $ X_{1} $ and $ X_{2} $ are independent. But, the Mellin transform can be used as $ \mathbb{E}\left[Z^{n}\right] = \mathcal{M}\left[f_{Z}\left(z\right);s\right]\vert_{s = n+1} $. Hence, from (\ref{Eq: ProductKMS_MellinTransRatio}) we have 
	\begin{equation}\label{Eq: ProductAKMS_RatioMoments}
		\begin{aligned}
			\mathbb{E}\left[Z^{n}\right] &= \theta_{1}\theta_{2} \left(\bar{\gamma}_{1}\right)^{n}\left(\bar{\gamma}_{2}\right)^{-n} c_{1}^{\frac{2n}{\alpha_{1}}} c_{2}^{\frac{-2n}{\alpha_{2}}}  \Gamma\left(\mu_{1} + \frac{2n}{\alpha_{1}}\right) \\& \times \Gamma\left(\mu_{2} - \frac{2n}{\alpha_{2}}\right)  {}_{2}F_{1}\left(m_{1},\mu_{1} + \frac{2n}{\alpha_{1}};\mu_{1};\beta_{1}\right) \\
			&\times {}_{2}F_{1}\left(m_{2},\mu_{2} - \frac{2n}{\alpha_{2}};\mu_{2};\beta_{2}\right).
		\end{aligned}
	\end{equation}
	\section{Integral representation}\label{Sec: ProductAKMS_IntegralRep}
	In the previous sections, we have derived the expressions for the PDF of the product and ratio statistics of two i.n.i.d. $ \alpha-\kappa-\mu $ shadowed RVs. However, the expression in \eqref{Eq: ProductAKMS_ExactProdPDF}, and \eqref{Eq: ProductAKMS_ExactRatioPDF} involves the H-function, which is not implemented in popular software like \textit{MATLAB}. The H-function is available in \textit{Mathematica} and can be used to evaluate the derived expression. Still, due to the presence of double infinite summation and considering the general nature of the values of the parameter involved, it is difficult to predict the number of terms to add for a particular level of accuracy.
	\par In this section, we have given the Laplace type integral representation for the derived expression using the method of residue and the definition of the Gamma function. These real-integral type expressions are far easier to evaluate using the Gauss-Legendre quadrature rule than evaluating a double infinite series with an H-function.
	\subsection{Laplace type integral for the PDF of the product statistics}
	By substituting the definition of H-function in \eqref{Eq: ProductAKMS_ExactProdPDF}, we have 
	\begin{equation}
		\begin{aligned}
			f_{Y}\left(y\right) &= \delta\theta_{1}\theta_{2}   \sum_{u,v=0}^{\infty}  \frac{A_{u,v}}{2\pi i} \int_{\mathcal{C}} \Gamma\left(\mu_{1}+u-\frac{2}{\alpha_{1}}+ \frac{2}{\alpha_{1}}s \right) \\& \times  \Gamma\left(\mu_{2} + v - \frac{2}{\alpha_{2}} + \frac{2}{\alpha_{2}}s \right) \left(\delta y\right)^{-s} \operatorname{ds}
		\end{aligned}
	\end{equation}
	Let $ \mu_{1}+u-\frac{2}{\alpha_{1}}+ \frac{2}{\alpha_{1}}s = \omega $ then, $ f_{Y}\left(y\right) $  can be written as
	\begin{equation}
		\begin{aligned}
			f_{Y}\left(y\right) &= \frac{\alpha_{1} \theta_{1}\theta_{2} \delta \left(\beta_{3}\right)^{\mu_{1}- \frac{2}{\alpha_{1}}}}{2} \sum_{u,v=0}^{\infty}  \frac{A_{u,v} \beta_{3}^{u}}{2\pi i} \int_{\mathcal{C}} \Gamma\left(\omega\right) \\&\times \Gamma\left(\mu_{2} - \frac{\alpha_{1}}{\alpha_{2}}\mu_{1} - \frac{\alpha_{1}}{\alpha_{2}}u  + v + \frac{\alpha_{1}}{\alpha_{2}}\omega \right) \beta_{3}^{-\omega} \operatorname{d\omega},
		\end{aligned}
	\end{equation}
	where $ \beta_{3} = \left(\delta y\right)^{\frac{\alpha_{1}}{2}} $. From \cite[Eq. 4]{kilbas2010kratzel} the Mellin-Barnes integral in the above equation can be identified as the Kr{\"a}tzel function. Hence, we have
	\begin{equation}
		\begin{aligned}
			f_{Y}\left(y\right) &= \frac{\alpha_{2} \theta_{1}\theta_{2} \delta \left(\beta_{3}\right)^{\mu_{1}- \frac{2}{\alpha_{1}}}}{2} \sum_{u,v=0}^{\infty}A_{u,v} \beta_{3}^{u} Z_{\rho}^{ \nu}\left(\beta_{3}\right),
		\end{aligned}
	\end{equation} 
	where $ \rho = \frac{\alpha_{2}}{\alpha_{1}} $, $ \nu = \frac{\alpha_{2}}{\alpha_{1}}\left(\mu_{2}+v\right) - \mu_{1} - u  $ and $ Z_{\rho}^{\nu}\left(\cdot \right) $ is the Kr{\"a}tzel function. Next, using the definition on Kr{\"a}tzel function \cite[Eq. 1]{kilbas2010kratzel} and some algebraic manipulation, a single Laplace type integral representation of $ f_{Y}\left(y\right) $ is 
	\begin{equation}\label{Eq: ProductAKMS_ProdPDF_IntegralForm}
		\begin{aligned}
			f_{Y}\left(y\right) &=  \frac{\alpha_{1} \theta_{1}\theta_{2} \delta \left(\beta_{3}\right)^{\mu_{1}- \frac{2}{\alpha_{1}}}}{2}  \int_{0}^{\infty} t^{\left(\mu_{2} - \frac{\alpha_{1}}{\alpha_{2}}\mu_{1}-1\right)} \Exp{-t} \\& \times {}_{1}F_{1}\left(m_{1};\mu_{1};t^{-\frac{\alpha_{1}}{\alpha_{2}}}\beta_{1}\beta_{3} \right) {}_{1}F_{1}\left(m_{2};\mu_{2};t\beta_{2}\right) \\&\times \Exp{-\beta_{3}t^{-\frac{\alpha_{1}}{\alpha_{2}}}} \operatorname{dt}.
		\end{aligned}
	\end{equation}
	\subsection{Laplace type integral for the PDF of the ratio statistics}
	Using the calculus of residues, the series form of PDF in \eqref{Eq: ProductAKMS_ExactRatioPDF} is as follows
	\begin{equation}
		\begin{aligned}
			f_{Z}\left(z\right) &= \begin{cases*}
				\frac{\alpha_{1}\zeta\theta_{1}\theta_{2}}{2}  \sum\limits_{u,v,n=0}^{\infty}  \frac{A_{u,v}\left( -1\right)^{n}}{n!} \\ \times \Gamma\left(\mu_{2} + \frac{\alpha_{1}}{\alpha_{2}}\mu_{1} + \frac{\alpha_{1}}{\alpha_{2}}u  + v + \frac{\alpha_{1}}{\alpha_{2}}n\right) \left(\zeta z\right)^{\frac{\alpha_{1}\left(b_{1}+n\right)}{2}} \\ \hspace{12em}   \text{for} \ \zeta z < 1 \\
				\frac{\alpha_{2}\zeta\theta_{1}\theta_{2}  }{2}  \sum\limits_{u,v,n=0}^{\infty}\frac{A_{u,v} \left( -1\right)^{n}}{n!} \\ \times \Gamma\left(\mu_{1} + \frac{\alpha_{2}}{\alpha_{1}}\mu_{2} + u + \frac{\alpha_{2}}{\alpha_{1}}v  + \frac{\alpha_{2}}{\alpha_{1}}n  \right) \left(\frac{1}{\zeta z}\right)^{\frac{\alpha_{2}\left(1 - a_{1}+n\right)}{2}} \\ \hspace{12em} \text{for} \ \zeta z > 1
			\end{cases*}
		\end{aligned}
	\end{equation}
	Next, using the definition of the Gamma function and some algebraic manipulations, the Laplace type integral for the ratio statistics is given by \eqref{Eq: ProductAKMS_RatioPDF_IntegralForm} on the top of the next page.
	\begin{figure*}[t!]
		\begin{equation}\label{Eq: ProductAKMS_RatioPDF_IntegralForm}
			\begin{aligned}
				f_{Z}\left(z\right) &= \begin{cases*}
					\frac{\alpha_{1}\zeta\theta_{1}\theta_{2} \left(\beta_{5}\right)^{\mu_{1}- \frac{2}{\alpha_{1}}}  }{2} \int\limits_{0}^{\infty}  t^{\left(\mu_{2} + \frac{\alpha_{1}}{\alpha_{2}}\mu_{1}-1\right)} \Exp{-t} {}_{1}F_{1}\left(m_{1};\mu_{1};t^{\frac{\alpha_{1}}{\alpha_{2}}}\beta_{1}\beta_{5} \right)  {}_{1}F_{1}\left(m_{2};\mu_{2};t\beta_{2}\right) \Exp{-\beta_{5}t^{\frac{\alpha_{1}}{\alpha_{2}}}}  \operatorname{dt} \ \text{for} \ \zeta z < 1, \\
					\frac{\alpha_{2}\zeta\theta_{1}\theta_{2}  \left(\beta_{6}\right)^{\mu_{2}+ \frac{2}{\alpha_{2}}} }{2}  \int\limits_{0}^{\infty}  t^{\left(\mu_{1} + \frac{\alpha_{2}}{\alpha_{1}}\mu_{2}-1\right)} \Exp{-t} {}_{1}F_{1}\left(m_{1};\mu_{1};t\beta_{1} \right) {}_{1}F_{1}\left(m_{2};\mu_{2};t^{\frac{\alpha_{2}}{\alpha_{1}}}\beta_{2}\beta_{6}\right) \Exp{-\beta_{6}t^{\frac{\alpha_{2}}{\alpha_{1}}}}  \operatorname{dt} \ \text{for} \ \zeta z > 1,
				\end{cases*}
			\end{aligned}
		\end{equation} 
		where $ \beta_{5} = \left(\zeta z\right)^{\frac{\alpha_{1}}{2}} $, and $ \beta_{6} = \left(\zeta z\right)^{-\frac{\alpha_{2}}{2}} $.
		\vspace{0.5em}
		\hrule
	\end{figure*}
	Note that ratio RV can be used to model the signal-to-interference ratio between the desired user and the interference user. Since both the users are present in the same propagation environment, it is practical to assume $ \alpha_{1} = \alpha_{2} = \alpha $. For this special case, it is easy to show that $  H_{1,1}^{1,1}\left[\zeta z \  \begin{array}{|c}
		\left(a_{1},\frac{2}{\alpha}\right)\\[2mm]
		\left(b_{1},\frac{2}{\alpha}\right)
	\end{array}  \right] = \frac{\alpha \Gamma\left(1-a_{1} + b_{1}\right)}{2} \frac{\left(\zeta z\right)^{\frac{\alpha b_{1}}{2}}}{\left(1 + \left(\zeta z\right)^{\frac{\alpha}{2}}\right)^{1-a_{1} + b_{1}}}  $. After substituting this in \eqref{Eq: ProductAKMS_ExactRatioPDF} and some algebraic manipulations, we have 
	\begin{equation}
		\begin{aligned}
			&f_{Z}\left(z\right)	= \frac{\alpha\zeta\theta_{1}\theta_{2}  \left(\beta_{7}\right)^{\mu_{1} - \frac{2}{\alpha}} }{2 \left(1 + \beta_{7}\right)^{\mu_{1} + \mu_{2}} }  \int\limits_{0}^{\infty} t^{\left(\mu_{1} + \mu_{2}-1\right)} \Exp{-t} \\& \times {}_{1}F_{1}\left(m_{1};\mu_{1};\frac{t\beta_{1} \beta_{7}}{1 + \beta_{7}} \right) {}_{1}F_{1}\left(m_{2};\mu_{2};\frac{t\beta_{2}}{1 + \beta_{7}} \right) \operatorname{dt},
		\end{aligned}
	\end{equation}
	where $\beta_{7} = \left(\zeta z \right)^{\frac{\alpha}{2}}$. The above integral can be identified as a double hypergeometric series (Appell series) $ F_{2} $ \cite[1.3.3]{srivastava1985:multiple}, and we have
	\begin{equation}\label{Eq: ProductAKMS_RatioPDF_IntegralFormSameAlpha}
		\begin{aligned}
			&f_{Z}\left(z\right)	= \frac{\alpha\zeta\theta_{1}\theta_{2}  \Gamma\left(\mu_{1}+\mu_{2}\right) \left(\beta_{7}\right)^{\mu_{1} - \frac{2}{\alpha}} }{2 \left(1 + \beta_{7}\right)^{\mu_{1} + \mu_{2}} } \\& \times F_{2}\left(\mu_{1}+\mu_{2},m_{1},m_{2};\mu_{1},\mu_{2};\frac{\beta_{1} \beta_{7}}{1 + \beta_{7}},\frac{\beta_{2}}{1 + \beta_{7}}\right)
		\end{aligned}
	\end{equation}
	\section{Asymptotic Analysis}
	In previous section, we derived the exact Laplace type integral expressions for the PDF of the product and ratio statistics of two i.n.i.d. $\alpha - \kappa - \mu $ shadowed RVs. In order to get more insight and simplified expression, we focused on the asymptotic analysis in this section.
	\subsection{For the Product Statistics}
	\par In high SNR regime, when either or both $\bar{\gamma}_{1},\bar{\gamma}_{2} \rightarrow \infty$. We have $\beta_{3} \rightarrow 0$. The PDF of the product statistics in \eqref{Eq: ProductAKMS_ProdPDF_IntegralForm} simplifies to 
	\begin{equation}
		\begin{aligned}
			f_{Y} \left(y\right) \mid_{\Uparrow}  &\approx \frac{\alpha_{1} \theta_{1}\theta_{2} \delta \left(\beta_{3}\right)^{\mu_{1}- \frac{2}{\alpha_{1}}}}{2}  \int_{0}^{\infty} t^{\left(\mu_{2} - \frac{\alpha_{1}}{\alpha_{2}}\mu_{1}-1\right)} \Exp{-t} \\& \times  {}_{1}F_{1}\left(m_{2};\mu_{2};t\beta_{2}\right)  \operatorname{dt}.  
		\end{aligned}
	\end{equation}     
	Using the identity \cite[7.621.4]{Grad2007}, we have 
	\begin{equation}
		\begin{aligned}
			f_{Y} \left(y\right) \mid_{\Uparrow}  &\approx \frac{\alpha_{1} \theta_{1}\theta_{2} \delta \left(\beta_{3}\right)^{\mu_{1}- \frac{2}{\alpha_{1}}}}{2} \Gamma\left(\mu_{2} - \frac{\alpha_{1}}{\alpha_{2}}\mu_{1}\right) \\& \times  {}_{2}F_{1}\left(m_{2}, \mu_{2} - \frac{\alpha_{1}}{\alpha_{2}}\mu_{1};\mu_{2};\beta_{2}\right).  
		\end{aligned}
	\end{equation}
	Hence, in high SNR regime, asymptotic CDF of the product statistics can be approximated as 
	\begin{equation}
		\begin{aligned}
			F_{Y} \left(y\right) \mid_{\Uparrow}  &\approx  \frac{\theta_{1}\theta_{2}}{\mu_{1}}   \Gamma\left(\mu_{2} - \frac{\alpha_{1}}{\alpha_{2}}\mu_{1}\right) \\& \times  {}_{2}F_{1}\left(m_{2}, \mu_{2} - \frac{\alpha_{1}}{\alpha_{2}}\mu_{1};\mu_{2};\beta_{2}\right) \beta_{3}^{\mu_{1}}.
		\end{aligned}
	\end{equation}
	\subsection{For the Ratio Statistics}
	When $ \bar{\gamma}_{1} \rightarrow \infty $, we have $ \beta_{5} \rightarrow 0 $. The PDF of the ratio statistics in \eqref{Eq: ProductAKMS_RatioPDF_IntegralForm} simplifies to 
	\begin{equation}
		\begin{aligned}
			f_{Z}\left(z\right) \mid_{\Uparrow} &= \frac{\alpha_{1}\theta_{1}\theta_{2} \zeta\left(\beta_{5}\right)^{\mu_{1}- \frac{2}{\alpha_{1}}}  }{2} \int\limits_{0}^{\infty}  t^{\left(\mu_{2} + \frac{\alpha_{1}}{\alpha_{2}}\mu_{1}-1\right)} \Exp{-t} \\
			&\times   {}_{1}F_{1}\left(m_{2};\mu_{2};t\beta_{2}\right)   \operatorname{dt}, \\
			&= \frac{\alpha_{1}\theta_{1}\theta_{2} \zeta\left(\beta_{5}\right)^{\mu_{1}- \frac{2}{\alpha_{1}}}  }{2} \Gamma\left(\mu_{2} + \frac{\alpha_{1}}{\alpha_{2}}\mu_{1}\right) \\ &\times {}_{2}F_{1}\left(m_{2},\mu_{2} + \frac{\alpha_{1}}{\alpha_{2}}\mu_{1}; \mu_{2}; \beta_{2} \right) 
		\end{aligned}
	\end{equation}
	Hence, the asymptotic approximation for the CDF of the ratio statistics is 
	\begin{equation}
		\begin{aligned}
			F_{Z}\left(z\right) \mid_{\Uparrow} &= \frac{\theta_{1}\theta_{2}}{\mu_{1}} \Gamma\left(\mu_{2} + \frac{\alpha_{1}}{\alpha_{2}}\mu_{1}\right) \\
			&\times {}_{2}F_{1}\left(m_{2},\mu_{2} + \frac{\alpha_{1}}{\alpha_{2}}\mu_{1}; \mu_{2}; \beta_{2} \right) \beta_{5}^{\mu_{1}}
		\end{aligned}
	\end{equation} 
	\section{Simple Approximation}
	Although the expressions, derived for the PDF of product and ratio statistics of two $\alpha-\kappa-\mu$ shadowed RVs in \eqref{Eq: ProductAKMS_ProdPDF_IntegralForm} and \eqref{Eq: ProductAKMS_RatioPDF_IntegralForm}, are exact and easy to evaluate numerically. For some practical applications, having a simpler approximation for the PDF is helpful, as it helps to analyze and make inferences for complex systems. Hence, we proposed the Gamma distribution as a closed-form approximation for the product and the Beta prime distribution as a closed-form approximation for the ratio of two $\alpha-\kappa-\mu$ shadowed RVs. The parameters of respective distributions can be obtained using the method of moments.  
	\subsection{Gamma Approximation for Product Statistics}
	We have $Y = X_{1} X_{2}$, where $X_{1}$ and $X_{2}$ follows $\alpha-\kappa-\mu$ shadowed distribution. The first two moments of $Y$ are evaluated using \eqref{Eq: ProductAKMS_ProductMoments} and are as follows 
	\begin{equation}
		\begin{aligned}
			\mathbb{E}\left[Y\right] &= \bar{\gamma}_{1}\bar{\gamma}_{2}, \\ 
			\mathbb{E}\left[Y^{2}\right] &= \theta_{1}\theta_{2} \left(\bar{\gamma}_{1}\bar{\gamma}_{2}\right)^{2} c_{1}^{\frac{4}{\alpha_{1}}} c_{2}^{\frac{4}{\alpha_{2}}}  \Gamma\left(\mu_{1} + \frac{4}{\alpha_{1}}\right) \\&\times \Gamma\left(\mu_{2} + \frac{4}{\alpha_{2}}\right)  {}_{2}F_{1}\left(m_{1},\mu_{1} + \frac{4}{\alpha_{1}};\mu_{1};\beta_{1}\right) \\
			&\times {}_{2}F_{1}\left(m_{2},\mu_{2} + \frac{4}{\alpha_{2}};\mu_{2};\beta_{2}\right)
		\end{aligned}
	\end{equation}  
	Let $ G\left(k,\theta\right)$ represents the Gamma distribution with shape parameter $k$ and scale parameter $\theta$. The PDF of $Y$ is approximated as
	\begin{equation}
		\begin{aligned}
			f_{Y}\left(y\right) = \frac{y^{k_{Y}-1} \Exp{-\frac{y}{\theta_{Y}}}}{\Gamma\left(k_{Y}\right)\theta_{Y}^{k_{Y}} } \qquad y > 0,			
		\end{aligned}
	\end{equation} 
	where $k, \theta$ are related to moments of $W$ in the following manner
	\begin{equation}
		\begin{aligned}
			k_{Y} = \frac{\mathbb{E}\left[Y\right]}{\theta}, \quad \theta_{Y} = \frac{\mathbb{V}\left[Y\right]}{\mathbb{E}\left[Y\right]}.
		\end{aligned}
	\end{equation} 
	The parameter for approximating Gamma distribution can be evaluated as follows
	\begin{equation}
		\begin{aligned}
			k_{Y} = \frac{1}{C_{pro}}, \quad \theta_{Y} = \bar{\gamma}_{1}\bar{\gamma}_{2} C_{pro},
		\end{aligned}
	\end{equation}
	where $ C_{pro} = \theta_{1}\theta_{2}  c_{1}^{\frac{4}{\alpha_{1}}} c_{2}^{\frac{4}{\alpha_{2}}}  \Gamma\left(\mu_{1} + \frac{4}{\alpha_{1}}\right)  \Gamma\left(\mu_{2} + \frac{4}{\alpha_{2}}\right) \allowbreak  {}_{2}F_{1}\left(m_{1},\mu_{1} + \frac{4}{\alpha_{1}};\mu_{1};\beta_{1}\right) {}_{2}F_{1}\left(m_{2},\mu_{2} + \frac{4}{\alpha_{2}};\mu_{2};\beta_{2}\right) - 1 $.
	\subsection{Beta prime Approximation for Ratio Statistics} 
\begin{figure}[!ht]
		\centering
		\includegraphics[width=0.45\textwidth]{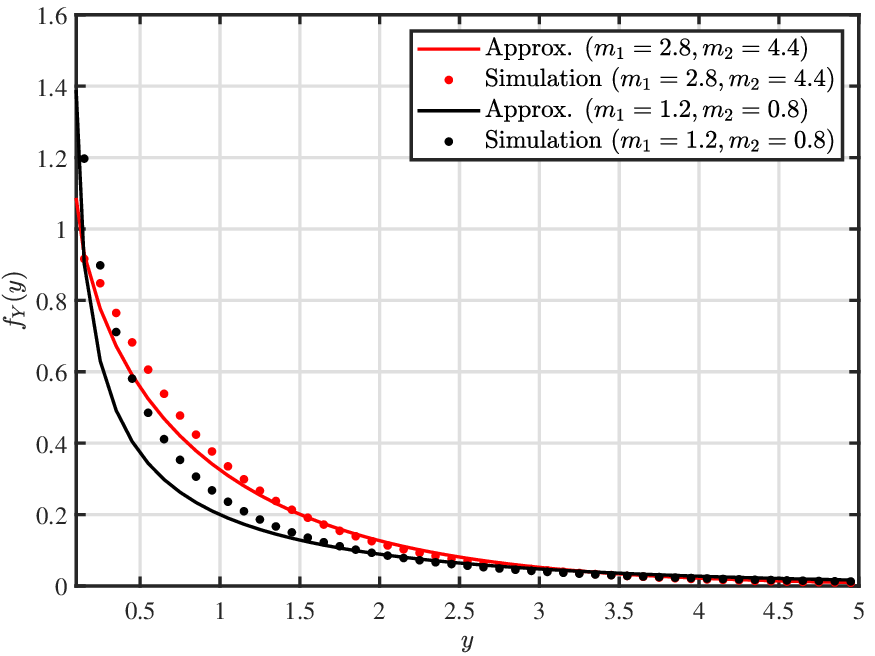}
		\captionof{figure}{Simulated and Approximate PDF of $Y$}
		\label{fig: ProdPDF_Approx}
	\end{figure}

\begin{figure}[!ht]
		\centering
		\includegraphics[width=0.45\textwidth]{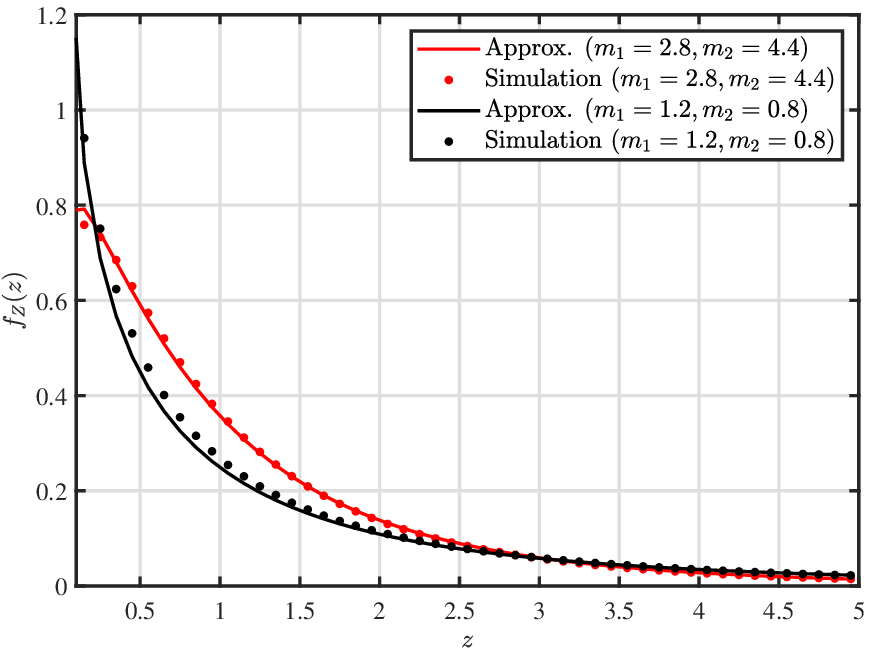}
		\captionof{figure}{Simulated and Approximate PDF of $Z$}
		\label{fig: RatioPDF_Approx}
	\end{figure}

	We have $Z = \frac{X_{1}}{X_{2}}$,  where $X_{1}$ and $X_{2}$ follows $\alpha-\kappa-\mu$ shadowed distribution. Similar to \cite{srinivasan2018secrecy}, we first approximated the $X_{1}$ and $X_{2}$ by a Gamma random variable using moment matching. Let $X_{1} \sim G \left(k_{Z1},\theta_{Z1}\right)$ and $X_{2} \sim G\left(k_{Z2},\theta_{Z2}\right)$, where $k_{Z1}, \theta_{Z1}, k_{Z2}, \theta_{Z2}$ are calculated as follows
	\begin{equation}
		\begin{aligned}
			k_{Z1} = \frac{1}{C_{ratio}}, \quad \theta_{Z1} = \bar{\gamma}_{1} C_{ratio},				
		\end{aligned}
	\end{equation}
	where $C_{ratio} = \theta_{1}  c_{1}^{\frac{4}{\alpha_{1}}}   \Gamma\left(\mu_{1} + \frac{4}{\alpha_{1}}\right)  {}_{2}F_{1}\left(m_{1},\mu_{1} + \frac{4}{\alpha_{1}};\mu_{1};\beta_{1}\right)  - 1$ and 
	\begin{equation}
		\begin{aligned}
			k_{Z2} = \frac{1}{D_{ratio}}, \quad \theta_{Z2} = \bar{\gamma}_{2} D_{ratio},
		\end{aligned}
	\end{equation}			 
	where $ D_{ratio} = \theta_{2}  c_{2}^{\frac{4}{\alpha_{2}}}   \Gamma\left(\mu_{2} + \frac{4}{\alpha_{2}}\right)  {}_{2}F_{1}\left(m_{2},\mu_{2} + \frac{4}{\alpha_{2}};\mu_{2};\beta_{2}\right)  - 1 $. Under the assumption that $X_{1}$ and $X_{2}$ follows Gamma distribution, $Z$ can be approximated by beta prime distribution \cite{cordeiro2012mcdonald}. The PDF of $Z$ is approximated as 
	\begin{equation}
		\begin{aligned}
			f_{Z}\left(z\right) &= \frac{\frac{\theta_{Z2}}{\theta_{Z1}} \left(\frac{\theta_{Z2}}{\theta_{Z1}} z\right)^{k_{Z1} - 1 }}{B\left(k_{Z1},k_{Z2}\right) \left(1 + \frac{\theta_{Z2}}{\theta_{Z1}} z\right)^{k_{Z1} + k_{Z2} } } \quad z > 0,
		\end{aligned}
	\end{equation}
	where $B\left(\cdot,\cdot\right)$ is the Beta function \cite{mathai2008special}.   
	Next, we have compared the proposed close-form approximations with simulated numerical values. In Fig. \ref{fig: ProdPDF_Approx} and \ref{fig: RatioPDF_Approx}, we have plotted the approximated and simulated PDF of $Y$ and $Z$, respectively for $ \alpha_{1} = 1.5, \kappa_{1} = 5.0, \mu_{1} = 1.2 $ and $ \alpha_{2} = 2.5, \kappa_{2} = 2.1, \mu_{2} = 3.0 $ with different values of $m_{1}$ and $m_{2}$.
	
	Fig. \ref{fig: ProdPDF_Approx} and \ref{fig: RatioPDF_Approx} shows that the proposed approximation are close to the simulated values of PDF and can be used in order to get a simplified expression in practical applications.
	\par In the next section, we discuss some applications where the product and ratio statistics of two $ \alpha-\kappa-\mu $ shadowed EV are useful. 	
	\section{Application Examples}\label{Sec: ProductAKMS_CommMetrics}
	\subsection{Cascaded Wireless System} Consider a two-tap cascaded channel  where both taps follow $\alpha-\kappa-\mu$
	shadowed fading. The considered model is a generalization of many existing works in literature such as \cite{bilim2022cascaded,Bhargav2018:ProductKappaMu,Silva:ProductAlphaKappaEtaMu}. This section derives the analytical expression for various important metrics
	for such a system.  
	\subsubsection{Outage Probability}
	In any communication system, an outage is when the signal's received strength falls below a certain threshold. The outage probability (OP) is defined as $ P_{OP}\left(\gamma_{th}\right) = \mathbb{P}\left(Y \le \gamma_{th}\right) $, hence, we have 
	\begin{equation}\label{Eq: ProductAKMS_CascadePout_1}
		\begin{aligned}
			P_{OP}\left(\gamma_{th}\right) &= \theta_{1}\theta_{2}  \sum_{u,v=0}^{\infty}A_{u,v} \\& \times H_{1,3}^{2,1}\left[\delta \gamma_{th} \  \begin{array}{|c}
				\left(0,1\right)\\
				\left(b_{3},\frac{2}{\alpha_{1}}\right), \left(b_{4},\frac{2}{\alpha_{2}}\right), \left(1,1\right)
			\end{array}  \right]
		\end{aligned}
	\end{equation}
	where $  b_{3} = \mu_{1}+u, b_{4} = \mu_{2}+v $. Apart from using this double-infinite series expression, we can also make use of \eqref{Eq: ProductAKMS_ProdPDF_IntegralForm} to compute the OP as $P_{OP}\left(\gamma_{th}\right) = \int\limits_{0}^{\gamma_{th}} f_{Y}\left(y \right) \operatorname{dy} $. This integral can be evaluated using the Gauss-Legendre quadrature rule \cite[25.4.29]{abramowitz1965:handbook}. 
	
	\subsubsection{Amount of Fading} The amount of fading (AF) measures the severity of any fading channel.
	It is defined as the ratio of variance to the square of the mean of instantaneous SNR \cite{simon2005digital}. Hence, the AF for cascaded $\alpha-\kappa-\mu$ shadowed channel is
	\begin{equation}\label{Eq:ProductAKMS_AF_Def}
		\begin{aligned}
			\text{AF} &= \frac{\mathbb{V}\left[Y\right]}{\left(\mathbb{E}\left[Y\right]\right)^{2}} = \frac{ \mathbb{E}\left[Y^{2}\right]-\left(\mathbb{E}\left[Y\right]\right)^{2}}{\left(\mathbb{E}\left[Y\right]\right)^{2}}.
		\end{aligned}
	\end{equation}
	The $\text{AF}$ can now be calculated using \eqref{Eq: ProductAKMS_ProductMoments}. 
	\par \textbf{Remark:} For $\alpha_{1} = \alpha_{2} = 2$ the result in \eqref{Eq: ProductAKMS_CascadePout_1} coincides with the corresponding result in \cite{bilim2022cascaded}. Similarly, for $\alpha_{i} = 2, m_{i} \rightarrow \infty$, our OP and AF expression reduces to the \cite[Eq. (20)]{Bhargav2018:ProductKappaMu} and \cite[Eq. (16)]{Bhargav2018:ProductKappaMu}, respectively. In \cite{Silva:ProductAlphaKappaEtaMu}, the cascade channel with $\alpha-\mu$, $\kappa-\mu$, and $\eta-\mu$ fading are considered since these three fading models are the special case of $\alpha-\kappa -\mu$ shadowed fading hence, the results encompass the corresponding results in \cite{Silva:ProductAlphaKappaEtaMu}.
	\subsection{Physical Layer Security}
	In \cite{bloch2008wireless,iwata2017secure,bhargav2016secrecy,srinivasan2018secrecy} instantaneous secrecy capacity of typical $3$ node single-input-single-output (SISO) wireless communication system is defined as 
	\begin{equation}
		\begin{aligned}
			C_{S} &= \left[\log_{2}\left(1 + \gamma_{SD} \right) - \log_{2}\left(1 + \gamma_{SE}\right)\right]^{+}
		\end{aligned}
	\end{equation} 
	where $ \left[x\right]^{+} = \max\curlybracket*{x,0} $. The $ \gamma_{SD} $ and $ \gamma_{SE} $ are instantaneous SNR over the source to legitimate destination and source to eavesdropper link, respectively. Both links are assumed to be experiencing $ \alpha - \kappa-\mu $ shadowed fading. The secrecy outage probability (SOP) is defined as the phenomenon in which the instantaneous secrecy capacity falls below a certain target secrecy rate $ R_{S} $, \textit{i.e.,} we have
	\begin{equation}
		\begin{aligned}
			P_{SOP}\left(R_{S}\right) &= \mathbb{P}\left[C_{S} \le R_{S}\right]
		\end{aligned}
	\end{equation}  
	Using \cite{belmoubarik2014secrecy}, $ P_{SOP}\left(R_{s}\right) $ is simplified as
	\begin{equation}
		\begin{aligned}
			P_{SOP}\left(R_{s}\right) &= \mathbb{P}\left[\frac{1 + \gamma_{SD}}{1 + \gamma_{SE}} \le \gamma_{th}\right]
		\end{aligned}
	\end{equation} 
	where $ \gamma_{th} = 2^{R_{s}} $. Applying the approximation $ \frac{1 + x}{1 + y} \approx \frac{x}{y} $ \cite{fan2014secure}, we have 
	\begin{equation}
		\begin{aligned}
			P_{SOP}\left(R_{s}\right) &\approx \mathbb{P}\left[\frac{\gamma_{SD}}{\gamma_{SE}} \le \gamma_{th}\right] 
		\end{aligned}
	\end{equation} 
	The above equation is nothing but the CDF of the ratio of two $ \alpha - \kappa-\mu $ shadowed RVs, that is given in \eqref{Eq: ProductAKMS_ExactRatioCDF}. Another secrecy performance metric is strictly positive secrecy capacity (SPSC) which is defined as $ P_{0} = \mathbb{P}\left[ C_{s} > 0\right] $. In terms of SOP, we can write SPSC as $ P_{0} = 1 - P_{SOP}\left(0\right)$.  
	\par \textbf{Remark: } Recently, the authors in \cite{badrudduza2021secrecy} studied the secrecy performance under $\alpha-\kappa-\mu$ shadowed fading, but the results are valid only for integer values of $\mu_{1}$ and $\mu_{2}$. Whereas the results derived in this paper have no such restriction. Since the $\alpha-\kappa-\mu$ shadowed fading is the general case for other fading scenarios such as Nakagami-$m$, $\kappa-\mu$, $\kappa-\mu$ shadowed hence, our results for SOP and SPSC encompasses the results in \cite{belmoubarik2014secrecy,iwata2017secure,bhargav2016secrecy,srinivasan2018secrecy}.

	\subsection{IRS-Assisted Communication System}\label{SubSec: ProductAKMS_SISO_IRS}
	Considering a system model similar to \cite{Charishma2021OutageIRS}, we have one single antenna source node ($\mathbf{S}$) communicating with one single antenna destination node ($\mathbf{D}$) using an  $\mathbf{IRS}$ with $N$ reflector elements. Let, $\mathbf{h}^{SR} \in  \mathbb{C}^{N\times1}$, $\mathbf{h}^{RD} \in \mathbb{C}^{N\times1}$ and ${h}^{SD} \in  \mathbb{C}^{1}$ denote the small-scale fading channel coefficients of the $\mathbf{S}$ to $\mathbf{IRS}$ (SR), $\mathbf{IRS}$ to $\mathbf{D}$ (RD) and $\mathbf{S}$ to $\mathbf{D}$ (SD) link, respectively. All the channels are assumed to experience independent $\alpha-\kappa-\mu$ shadowed fading \textit{i.e.}, each of $\abs*{\left[\mathbf{h}^{SR}\right]_{i}}^{2}$, $\abs*{\left[\mathbf{h}^{RD}\right]_{i}}^{2}$ and $ \abs*{h^{SD}}^{2} $ has the PDF as given in \eqref{Eq: ProductAKMS_AKappaMuSpdf} with $ \mathbb{E}\left[\abs*{\left[\mathbf{h}^{AB}\right]_{i}}^{2}\right] = \gamma_{AB} = d_{AB}^{-\beta}$ where $A,B \in \curlybracket*{S,R,D}$. The $d_{AB}$ represents the distance between node $A, B$, and $\beta$ is the path-loss coefficient. Let $\theta_{i} = \text{arg}\left({h}^{SD}\right)- \text{arg}\left(\left[\mathbf{h}^{SR}\right]_{i}\left[\mathbf{h}^{RD}\right]_{i}\right)$ be the phase shift introduced by the $i$-{th} $\mathbf{IRS}$ element. This phase-shift design is widely used in IRS literature, such as \cite{bjornson2019intelligent,li2020ergodic,de2021large, Charishma2021OutageIRS}.   Then the SNR at the node $\mathbf{D}$ is 
	\begin{equation}\label{Eq: ProductAKMS_IRS_snr_err}
		\begin{aligned}
			\gamma_{IRS} =\gamma_{s} \abs*{ \abs*{{h}^{SD}} + \sum\limits_{i=1}^{N} \abs*{\left[\mathbf{h}^{SR}\right]_{i}} \abs*{\left[\mathbf{h}^{RD}\right]_{i}} }^{2},
		\end{aligned}
	\end{equation}
	where $\gamma_{s}$ is the ratio of transmitted power $p$ and receiver noise power $ \sigma^{2}$. In any communication system, an outage is when the signal's received strength falls below a certain threshold. Let  $\abs{{h}^{SD}}, \abs{\left[\mathbf{h}^{SR}\right]_{i}} $ and $\abs{\left[\mathbf{h}^{RD}\right]_{i}}$ be denoted by $g_{SD}, \left[\mathbf{g}_{SR}\right]_{i} $ and $ \left[\mathbf{g}_{RD}\right]_{i}$ respectively. The outage probability (OP) for threshold $\gamma_{th}$ can be evaluated as
	\begin{equation}\label{Eq:LIS_CLT1}
		\begin{aligned}
			P_{out}\left( \gamma_{th} \right) & = \mathbb{P}\left( \gamma_{s} \abs*{ \left( {g}^{SD} +  \underbrace{\sum\limits_{i=1}^N [\mathbf{g}^{SR}]_{i} [\mathbf{g}^{RD}]_{i}}_{u} \right)}^{2} \le \gamma_{th} \right) \\&= \mathbb{P}\left(z \leq \sqrt{\frac{\gamma_{th}}{\gamma_{s}}}\right),
		\end{aligned}
	\end{equation}
	where $z \triangleq u + g^{SD} $. Now approximate the $u$ by the Gamma RV, \textit{i.e.,} $u \sim \operatorname{G}\left(k_{mom},\theta_{mom} \right)$ with
	\begin{equation}
		\begin{aligned}
			k_{mom} = \frac{N \mu^{2} }{\sigma^{2}}, \quad \theta_{mom} = \frac{\sigma^{2}}{\mu},
		\end{aligned}
	\end{equation}
	where $\mu$, and $\sigma^{2}$ are evaluated using \eqref{Eq: ProductAKMS_ProductMoments} as follows
	\begin{equation}
		\begin{aligned}
			\mu &= \theta_{SR}\theta_{RD} \left(\bar{\gamma}_{SR}\bar{\gamma}_{RD}\right)^{0.5} c_{SR}^{\frac{1}{\alpha_{SR}}} c_{RD}^{\frac{1}{\alpha_{RD}}}  \Gamma\left(\mu_{SR} + \frac{1}{\alpha_{SR}}\right) \\& \times  \Gamma\left(\mu_{RD} + \frac{1}{\alpha_{RD}}\right)  {}_{2}F_{1}\left(m_{SR},\mu_{SR} + \frac{1}{\alpha_{SR}};\mu_{SR};\beta_{SR}\right) \\
			&\times {}_{2}F_{1}\left(m_{RD},\mu_{RD} + \frac{1}{\alpha_{RD}};\mu_{RD};\beta_{RD}\right),
		\end{aligned}
	\end{equation} 
	and the $\sigma^{2} = \bar{\gamma}_{SR}\bar{\gamma}_{RD} - \mu^{2}$. 
	The $z$ is a sum of two independent RV, so the CDF of the $z$ is as follows
	\begin{equation}\label{Eq:LIS_CLT2}
		F_{z}(z) = \int_{-\infty}^{\infty} f_{u}(z-x) F_{g^{SD}}(x) \operatorname{dx},
	\end{equation}
	where $f_{u}(x) = \frac{1}{\Gamma\left(k_{mom} \right) \theta_{mom}^{k_{mom}}} x^{k_{mom} - 1} \operatorname{e}^{-\frac{x}{\theta_{mom}}} $ is the pdf of $u$ and $F_{g^{SD}}(x)$ is the CDF of $g^{SD}$. After substituting values, we have
	\begin{equation}
		\begin{aligned}
			&F_{z}(z) = \frac{\theta_{SD}}{\mu_{SD} c_{SD}^{\mu_{SD}}\Gamma\left(k_{mom} \right) \theta_{mom}^{k_{mom}}} \int_{0}^{z}  \left(z-x\right)^{k_{mom} - 1}  \\
			&  \times \operatorname{e}^{-\frac{z-x}{\theta_{mom}}}  \left(\lambda\left(x \right)\right)^{\mu_{SD}} \\
			& \times \Phi_{2}\Bigg(\mu_{SD}-m_{SD},m_{SD};\mu_{SD} +  1; \Bigg. \\  &  \Bigg. \hspace{4em}\frac{-\lambda\left(x \right)}{c_{SD}}, -\lambda\left(x \right)\frac{1-\beta_{SD}}{c_{SD}} \Bigg)\operatorname{dx},
		\end{aligned}
	\end{equation}
	where $\Phi_{2}\left(\cdot,\cdot; \cdot; \cdot, \cdot \right)$ is the confluent bivariate hypergeometric function \cite{srivastava1985:multiple}, $  \lambda\left(x \right) = \left(\frac{x^{2}}{\bar{\gamma}_{SD} }\right)^{\frac{\alpha_{SD}}{2}} $. Let $ \frac{z-x}{\theta_{mom}} = t \Rightarrow \operatorname{d} x = - \theta_{mom} \operatorname{d} t$ then, OP is given by \eqref{Eq: ProductAKMS_PoutSISO_IRS_Gamma} on the top of next page.
	\begin{figure*}[t!]
		\begin{equation}\label{Eq: ProductAKMS_PoutSISO_IRS_Gamma}
			\begin{aligned}
				&P_{out}\left( \gamma_{th} \right) \approx \frac{\theta_{SD}}{\mu_{SD} c_{SD}^{\mu_{SD}}\Gamma\left(k_{mom} \right) } \int_{0}^{\frac{\sqrt{\frac{\gamma_{th}}{\gamma_{s}}}}{\theta_{mom}} } t^{k_{mom} - 1} \operatorname{e}^{-t}  \left(\lambda\left(\sqrt{\frac{\gamma_{th}}{\gamma_{s}}} - t \theta_{mom} \right)\right)^{\mu_{SD}} \\
				&  \times \Phi_{2}\left(\mu_{SD}-m_{SD},m_{SD};\mu_{SD} +  1; \frac{-\lambda\left(\sqrt{\frac{\gamma_{th}}{\gamma_{s}}} - t \theta_{mom} \right)}{c_{SD}}, -\lambda\left(\sqrt{\frac{\gamma_{th}}{\gamma_{s}}} - t \theta_{mom} \right)\frac{1-\beta_{SD}}{c_{SD}} \right)\operatorname{dt} 
			\end{aligned}
		\end{equation}
	\end{figure*}
	
	\par For the large values of $k_{mom}$, the expression in \eqref{Eq: ProductAKMS_PoutSISO_IRS_Gamma} can be numerically unstable due to the presence of $\Gamma\left(k_{mom}\right)$. In such case, we can use a widely known fact in statistics that, for large $k_{mom}$, a Gamma RV is approximated via Gaussian RV \cite{casella2021statistical}. Using the Gaussian approximation for $u$, OP for threshold $\gamma_{th}$ is given in \eqref{Eq: ProductAKMS_PoutSISO_IRS_Gaussian} on the top of next page,
	\begin{figure*}[t!]
		\begin{equation}\label{Eq: ProductAKMS_PoutSISO_IRS_Gaussian}
			\begin{aligned}
				P_{out}\left( \gamma_{th} \right) &\approx  \frac{\theta_{SD}}{\mu_{SD} c_{SD}^{\mu_{SD}}\sqrt{\pi}} \int_{\frac{ N\mu - \sqrt{\frac{\gamma_{th}}{\gamma_{s}}}}{\sqrt{2 N \sigma^{2}}}}^{\infty} \exp\left(- t^{2} \right) \left(h\left( t \right)\right)^{\mu_{SD}}  \\
				& \times \Phi_{2}\Bigg(\mu_{SD}-m_{SD},m_{SD};\mu_{SD} +  1; \frac{-h\left( t \right)}{c_{SD}}, -h\left( t \right)\frac{1-\beta_{SD}}{c_{SD}} \Bigg)\operatorname{dt}
			\end{aligned}
		\end{equation}
		\vspace{0.5em}
		\hrule
	\end{figure*}
	where $ h\left( t \right) = \lambda\left( \frac{t}{\sqrt{2 N \sigma^{2}}} - \frac{ N\mu - \sqrt{\frac{\gamma_{th}}{\gamma_{s}}}}{\sqrt{2 N \sigma^{2}}}\right) $. The utility of \eqref{Eq: ProductAKMS_PoutSISO_IRS_Gamma} and \eqref{Eq: ProductAKMS_PoutSISO_IRS_Gaussian} is demonstrated in Section \ref{Sec: ProductAKMS_Results}. Note that the confluent bivariate hypergeometric function $\Phi_{2}$ is not implemented in \textit{MATLAB}. However, a \textit{MATLAB} program for the evaluation of $\Phi_{2}$ is given in \cite{martos2016matlab} that is used in the evaluation of \eqref{Eq: ProductAKMS_PoutSISO_IRS_Gamma} and \eqref{Eq: ProductAKMS_PoutSISO_IRS_Gaussian}.
	\par \textbf{Remark: } The results derived here are quite general and encompass many previous results in the literature as special cases. For example, the \eqref{Eq: ProductAKMS_PoutSISO_IRS_Gaussian} reduces to \cite[Theorem 2]{tao2020performance} when SD link is Rayleigh distributed, SR and RD link are Rician distributed, \textit{i.e.} $\alpha_{SD} = 2, \kappa_{SD} = 0, \mu_{SD} = 1$, $m_{SD} \rightarrow \infty$ and $ \alpha_{SR} = \alpha_{RD} = 2, \kappa_{SR} = K_{1}, \kappa_{RD} = K_{2}, \mu_{SR} = \mu_{RD} = 1, m_{SR}, m_{RD} \rightarrow \infty $. Similarly, the OP expression given in \cite[Eq. (10)]{kudathanthirige_icc_20} and \cite[Eq. (15)]{atapattu2020reconfigurable}  are special cases when the SD link is in the outage, and the SR, RD link follows Rayleigh distribution. Similarly, the derived results generalize the work in \cite{selimis2021performance,Charishma2021OutageIRS} for no phase error case.
	
	\section{Numerical Results}\label{Sec: ProductAKMS_Results}
\begin{figure}[!ht]
		\centering
		\includegraphics[width=0.45\textwidth]{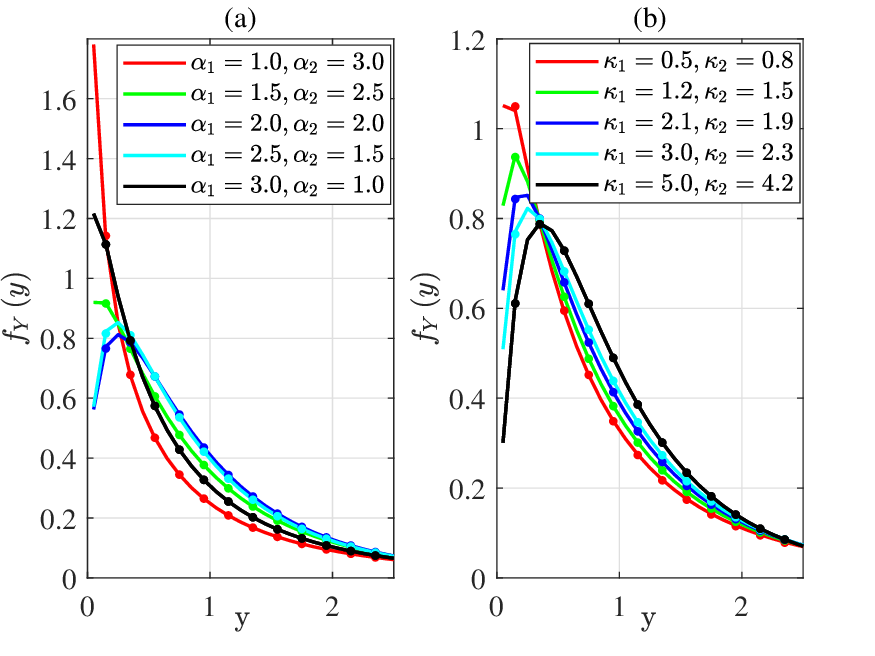}
		\captionof{figure}{PDF of $ Y $ for various values of $ \alpha_{1} $, $ \alpha_{2} $ in (a), and for various values of $ \kappa_{1} $, $ \kappa_{2} $ in (b).}
		\label{Fig: Prod_PDF_AKMS_Sim_Theo_various_Alpha_Kappa}
	\end{figure}

\begin{figure}[!ht]
		\centering
		\includegraphics[width=0.45\textwidth]{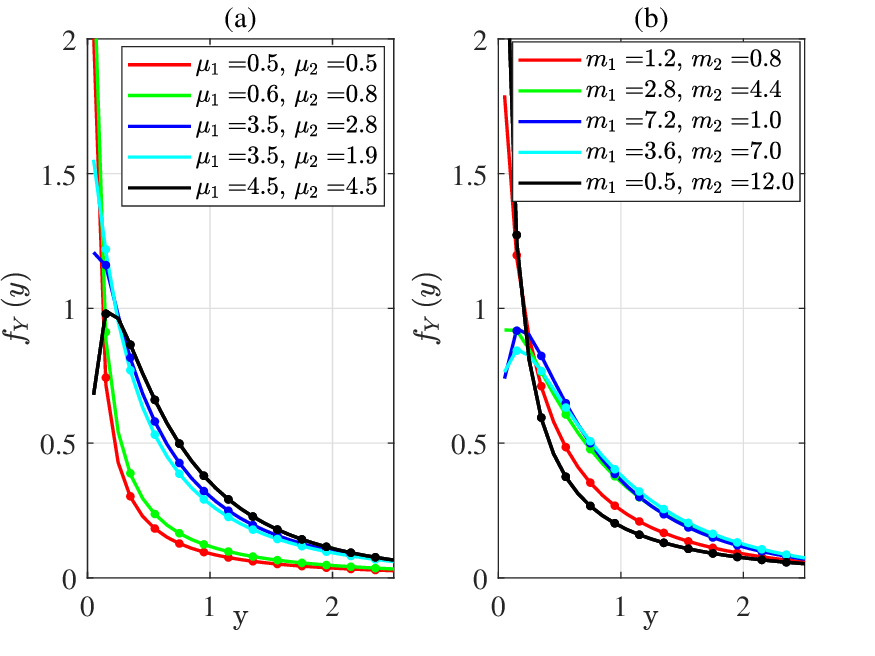}
		\captionof{figure}{PDF of $ Y $ for various values of $ \mu_{1} $, $ \mu_{2} $ in (a), and for various values of $ m_{1} $, $ m_{2} $ in (b).}
		\label{Fig: Prod_PDF_AKMS_Sim_Theo_various_Mu_M}
	\end{figure}

	This section presents the simulation results that show the correctness and utility of the theoretical expression presented in the previous sections.  Without loss of generality, we have assumed $ \bar{\gamma}_{1} = \bar{\gamma}_{2} = 1 $ for all the plots, unless mentioned otherwise. In all the figures, we have used solid lines to draw the theoretical values and the dotted markers are for simulated values. In Figs. \ref{Fig: Prod_PDF_AKMS_Sim_Theo_various_Alpha_Kappa} and \ref{Fig: Prod_PDF_AKMS_Sim_Theo_various_Mu_M}, several plots for the product PDF of two $ \alpha-\kappa-\mu $ shadowed RV for various values of $ \alpha, \kappa,\mu $ and $ m $ are given. In Fig. \ref{Fig: Prod_PDF_AKMS_Sim_Theo_various_Alpha_Kappa}(a), we have plotted PDF of product of two $ \alpha-\kappa-\mu $ shadowed RVs with $\kappa_{1} = 5.0, \mu_{1} = 1.2, m_{1} = 2.8$ and $\kappa_{2} = 2.1, \mu_{2} = 3.0, m_{2} = 4.4$ for different combination of $\alpha_{1}$ and $\alpha_{2}$ values. Similarly, in Fig. \ref{Fig: Prod_PDF_AKMS_Sim_Theo_various_Alpha_Kappa}(b), we have plotted the PDF of $Y$ for different combination of $\kappa_{1}$ and $\kappa_{2}$ values with $\alpha_{1} = 1.5, \mu_{1} = 2.1, m_{1} = 10.0$ and $\alpha_{2} = 2.5, \mu_{2} = 1.5, m_{2} = 4.0$.  
	Next, we have plotted the PDF of $Y$ for different combinations of $\mu$ and $m$ in Fig. \ref{Fig: Prod_PDF_AKMS_Sim_Theo_various_Mu_M}. Particularly, in Fig. \ref{Fig: Prod_PDF_AKMS_Sim_Theo_various_Mu_M}(a) for $\alpha_{1} = 1.0, \kappa_{1} = 2.2, m_{1} = 10.0$ and $\alpha_{2} = 1.5, \kappa_{2} = 0.9, m_{2} = 4.0$, the plots are given for different combination of $\mu_{1}$ and $\mu_{2}$. Lastly, Fig. \ref{Fig: Prod_PDF_AKMS_Sim_Theo_various_Mu_M}(b) has the results for different combination of $m_{1}$ and $m_{2}$ with $\alpha_{1} = 1.5, \kappa_{1} = 5.0, \mu_{1} = 1.2$ and $\alpha_{2} = 2.5, \kappa_{2} = 2.1, \mu_{2} = 3.0$. Hence, one can observe that a wide range of shapes can be fitted through double $ \alpha-\kappa-\mu $ shadowed RV. Also, note that the simulated PDFs are perfectly matching with the values obtained through theoretical expression in \eqref{Eq: ProductAKMS_ProdPDF_IntegralForm}, which shows the correctness and utility of expression.     
	
\begin{figure}[!ht]
		\centering
		\includegraphics[width=0.45\textwidth]{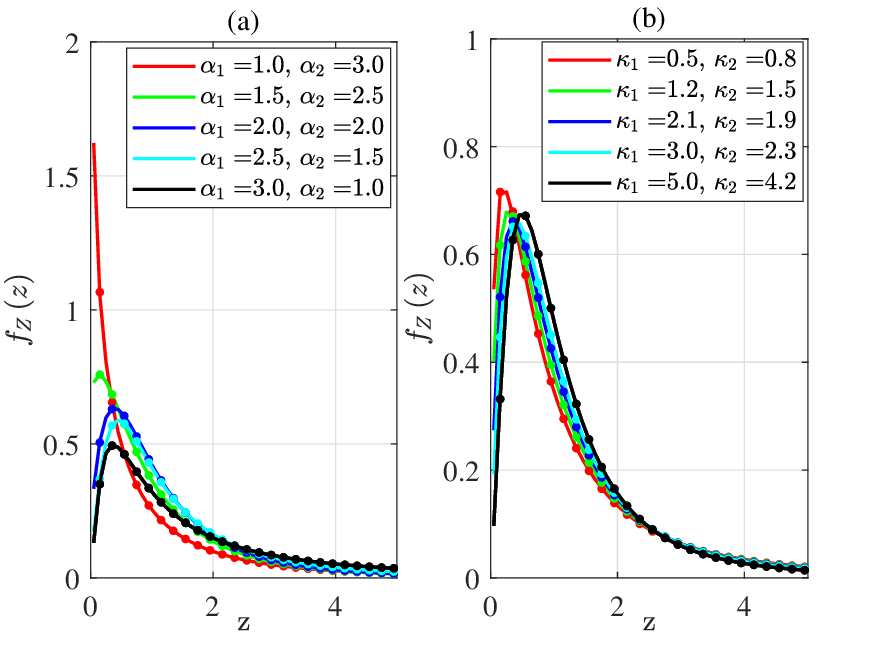}
		\captionof{figure}{PDF of $ Z $ for various values of $ \alpha_{1} $, $ \alpha_{2} $ in (a), and for various values of $ \kappa_{1} $, $ \kappa_{2} $ in (b).}
		\label{Fig: Ratio_PDF_AKMS_Sim_Theo_various_Alpha_Kappa}
	\end{figure}

\begin{figure}[!ht]
		\centering
		\includegraphics[width=0.45\textwidth]{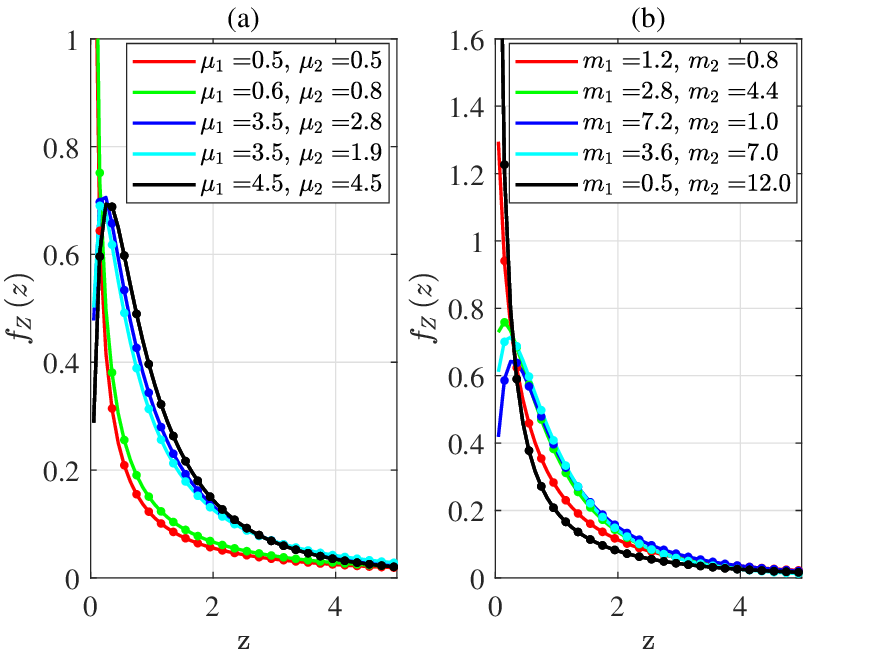}
		\captionof{figure}{PDF of $ Z $ for various values of $ \mu_{1} $, $ \mu_{2} $ in (a), and for various values of $ m_{1} $, $ m_{2} $ in (b).}
		\label{Fig: Ratio_PDF_AKMS_Sim_Theo_various_Mu_M}
	\end{figure}

	Next, We plotted the PDF values for the ratio of two $ \alpha-\kappa-\mu $ shadowed RVs in Figs.  
	\ref{Fig: Ratio_PDF_AKMS_Sim_Theo_various_Alpha_Kappa} and \ref{Fig: Ratio_PDF_AKMS_Sim_Theo_various_Mu_M}. The parameters values in Fig. \ref{Fig: Ratio_PDF_AKMS_Sim_Theo_various_Alpha_Kappa}(a) and (b) are the same as kept for Fig. \ref{Fig: Prod_PDF_AKMS_Sim_Theo_various_Alpha_Kappa}(a) and (b). Similar is the case with Fig. \ref{Fig: Ratio_PDF_AKMS_Sim_Theo_various_Mu_M}(a),(b) and Fig. \ref{Fig: Prod_PDF_AKMS_Sim_Theo_various_Mu_M}(a),(b). In all the plots, the value of PDF obtained through simulated RV and theoretical expression in \eqref{Eq: ProductAKMS_RatioPDF_IntegralForm} and \eqref{Eq: ProductAKMS_RatioPDF_IntegralFormSameAlpha}.

	In the following subsections, we have presented the simulation results for the applications discussed in Section \ref{Sec: ProductAKMS_CommMetrics}.
	
	\subsection{Results for Cascaded Wireless System}
\begin{figure}[!ht]
		\centering
		\includegraphics[width=0.45\textwidth]{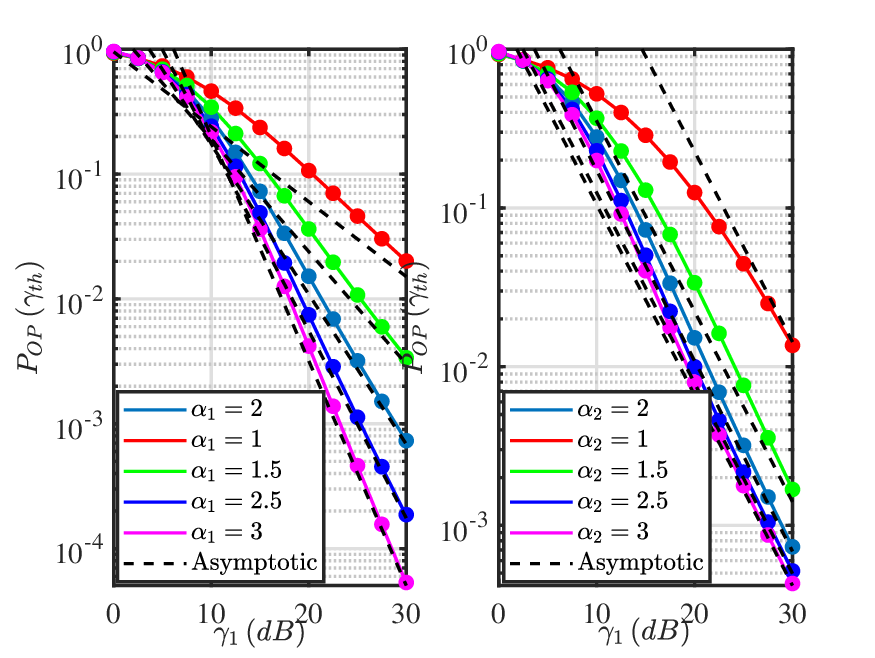}
		\captionof{figure}{OP of cascaded $\alpha-\kappa-\mu$ shadowed channel}
		\label{fig: OP_Cascade_AKMS_Vary_a}
	\end{figure}

\begin{figure}[!ht]
		\centering
		\includegraphics[width=0.45\textwidth]{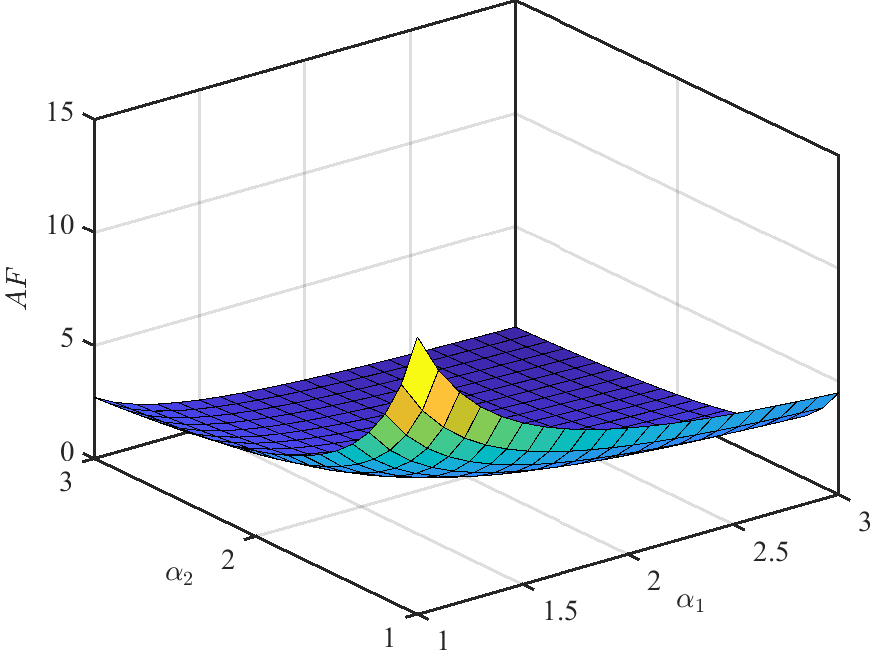}
		\captionof{figure}{AF of cascaded $\alpha-\kappa-\mu$ shadowed channel}
		\label{fig: AF_Cascade_AKMS_Vary_a}
	\end{figure}

	This subsection presents the OP and AF for a cascaded wireless system over $\alpha-\kappa-\mu$ shadowed fading. Fig. \ref{fig: OP_Cascade_AKMS_Vary_a} and \ref{fig: AF_Cascade_AKMS_Vary_a} show the impact of $\alpha$ parameter on OP and AF, respectively, while other parameters are kept constant.
	
	In Fig \ref{fig: OP_Cascade_AKMS_Vary_a}(a), we kept $ \kappa_{1} = 5.0, \mu_{1} = 1.2 $, $m_{1} = 3.6$ and $\alpha_{2} = 2.0, \kappa_{2} = 2.1, \mu_{2} = 3.0, m_{2} = 1.0 $ with $\bar{\gamma}_{2} = 0$ dB. Threshold SNR, \textit{i.e.,} $\gamma_{th}$ is set to be $5$ dB. The OP of the cascade system with varying values $\bar{\gamma}_{1} $ is plotted for different values of $\alpha_{1}$. Results demonstrate that the OP decreases as $\alpha_{1}$ increases. This trend can also be verified through AF values shown in Fig. \ref{fig: AF_Cascade_AKMS_Vary_a} where we can observe that the AF decreases as either $\alpha_{1}$ or $\alpha_{2}$ increases. Hence, the degrading effect of fading will be less for high values of $\alpha_{1}$ or $\alpha_{2}$. Similarly in Fig. \ref{fig: OP_Cascade_AKMS_Vary_a}(b) we kept $ \alpha_{1} = 2.0, \kappa_{1} = 5.0, \mu_{1} = 1.2, m_{1} = 3.6 $ and $ \kappa_{2} = 2.1, \mu_{2} = 3.0, m_{2} = 1.0 $ then we changed the value of $ \alpha_{2} $. It is observed that as $ \alpha $ increases, the OP decreases. However, the effect is not independent of other parameters as in Fig. \ref{fig: OP_Cascade_AKMS_Vary_a}(a) the impact of increasing $ \alpha $ is more dominant compared to Fig. \ref{fig: OP_Cascade_AKMS_Vary_a}(b) where it saturates for $ \alpha_{2} = 2.5 $. One reason for this behavior may be that the $ \kappa_{1} > \kappa_{2}$ dominates the overall link. These theoretical results can be further used in resource allocation optimization problems where the objective is to minimize the OP.

	\subsection{Results for Physical Layer Security}
\begin{figure}[!ht]
		\centering
		\includegraphics[width=0.45\textwidth]{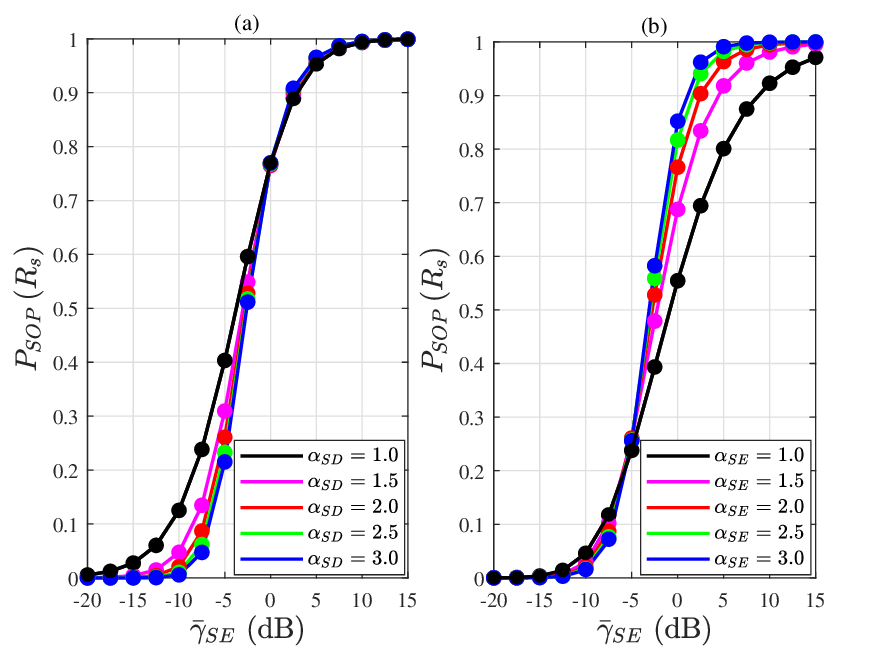}
		\captionof{figure}{SOP for $\alpha-\kappa-\mu$ shadowed fading for varying $\bar{\gamma}_{SE}$}
		\label{fig: SOP_AKMS_Vary_a}
	\end{figure}

\begin{figure}[!ht]
		\centering
		\includegraphics[width=0.45\textwidth]{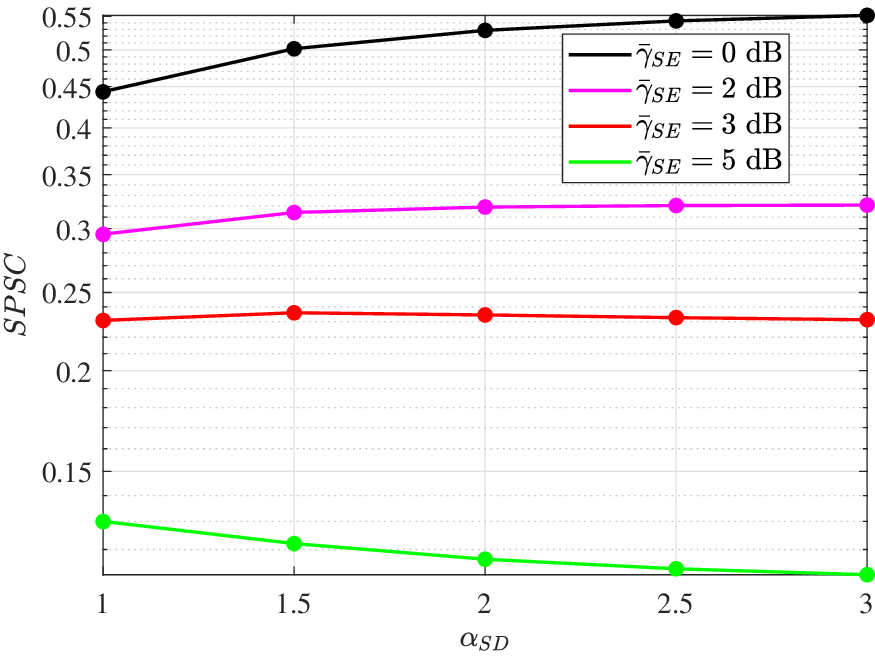}
		\captionof{figure}{SPSC for $\alpha-\kappa-\mu$ shadowed fading}
		\label{fig: SPSC_AKMS_Vary_a}
	\end{figure}

	This subsection presents the results for physical layer security metrics such as SOP and SPSC for $\alpha-\kappa-\mu$ shadowed fading. In Fig. \ref{fig: SOP_AKMS_Vary_a} and \ref{fig: SPSC_AKMS_Vary_a}, the fading parameters of legitimate (source to destination) link are $\kappa_{SD} = 5.0, \mu_{SD} = 2.1, m_{SD} = 10.0, \bar{\gamma}_{SD} = 0$ dB. Also, the fading parameters of the source to eavesdropper links are $\kappa_{SE} = 4.2, \mu_{SE} = 1.5, m_{SE} = 4.0$. The target secrecy rate is $R_{S} = 1$. In Fig. \ref{fig: SOP_AKMS_Vary_a}(a), we have plotted the SOP versus $\bar{\gamma}_{SE}$ for $\alpha_{SE} = 2.0$ and different values of $\alpha_{SD}$. Here, one interesting point to note is that a higher value of $\alpha_{SD}$ is not better for all the values of $\bar{\gamma}_{SE}$ as one can observe from the figure that a lower $\alpha_{SD}$ is having lower SOP for $\bar{\gamma}_{SE} > 0$  dB. Similarly, in Fig. \ref{fig: SOP_AKMS_Vary_a}(b), we have plotted the SOP versus $\bar{\gamma}_{SE}$ for $\alpha_{SD} = 2.0$ and different values of $\alpha_{SE}$.  Here, also we can see a crossover in the behavior of SOP at $\bar{\gamma}_{SE} = -5$ dB. Intuitively, an increase in $\alpha_{SE}$ should be detrimental to SOP but the result reveals that it is not independent from the value of $\bar{\gamma}_{SE}$. Increasing $\alpha_{SE}$ has detrimental effect only when the $\bar{\gamma}_{SE} \ge 0$ dB. 
	
	\begin{figure}[!ht]
		\centering
		\includegraphics[width=0.45\textwidth]{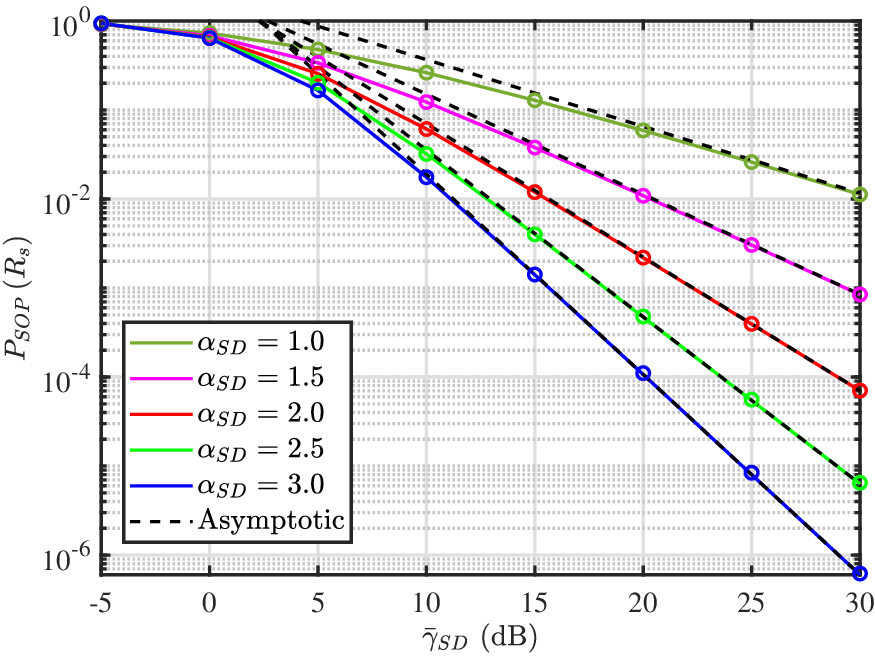}
		\captionof{figure}{SOP for $\alpha-\kappa-\mu$ shadowed fading for varying $\bar{\gamma}_{SD}$}
		\label{fig: SOP_AKMS_Vary_gSD}
	\end{figure}
	
	Fig. \ref{fig: SPSC_AKMS_Vary_a} present the SPSC versus $\alpha_{SD}$ for $\alpha_{SE} = 2.0$ and different values of $\bar{\gamma}_{SE}$. It is interesting to note that SPSC does not have a monotonic relation with $\alpha_{SD}$, \textit{i.e.,} we can't always expect a better SPSC for higher $\alpha_{SD}$. We can observe from the figure that for the high value of $\bar{\gamma}_{SE}$ SPSC saturates or even decreases with increasing $\alpha_{SD}$.
	\subsection{Results for IRS-Assisted Communication System}
\begin{figure}[!ht]
		\centering
		\includegraphics[width=0.45\textwidth]{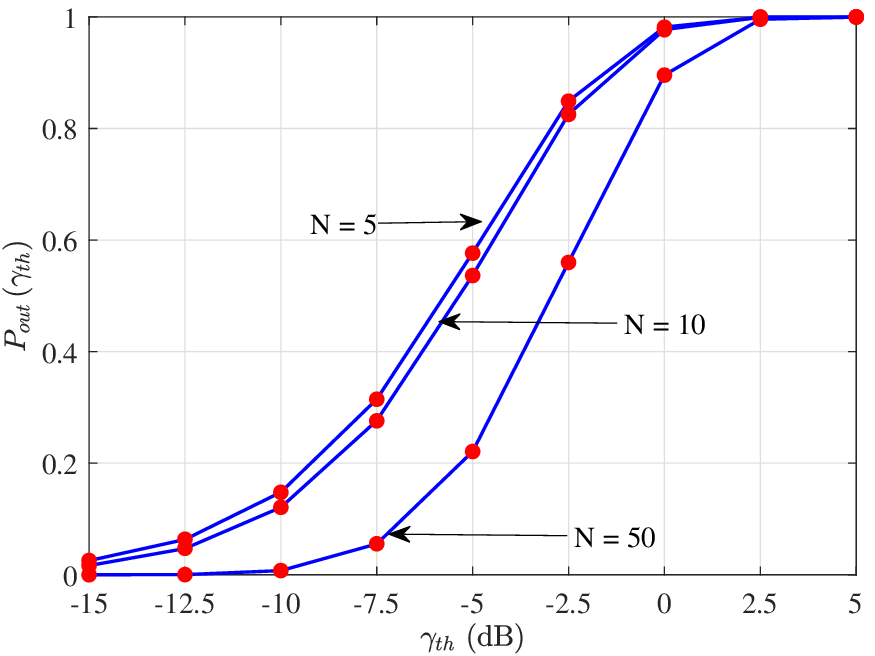}
		\captionof{figure}{OP of IRS-Assisted Communication System for various $N$}
		\label{fig: OP_SISO_IRS_AKMS_VaryN_Exp1}
	\end{figure}

\begin{figure}[!ht]
		\centering
		\includegraphics[width=0.45\textwidth]{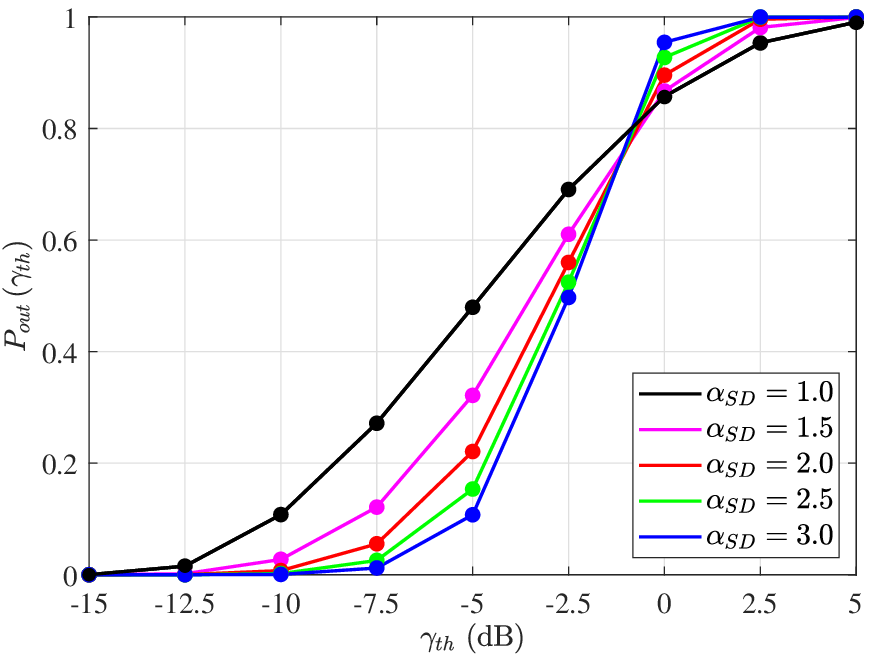}
		\captionof{figure}{OP of IRS-Assisted Communication System for various $\alpha_{SD}$}
		\label{fig: OP_SISO_IRS_AKMS_Vary_aSD_Exp1}
	\end{figure}

	In this subsection, we have presented the results of the IRS-assisted communication system. We consider a simulation setup such that the $\mathbf{S}$, IRS, and $\mathbf{D}$ are placed at $(0,0), (0,10)$, and $(90,0)$, respectively. The path-loss coefficient $\beta$ is chosen to be $4$ and $\gamma_{s} = 73$ dB. Other fading parameters are as follows  $ \kappa_{SD} = 0.8$, $ \mu_{SD} = 1.5$, $ m_{SD} = 4.0$,  $\kappa_{SR} = 2.1$, $\mu_{SR} = 3.0$, $m_{SR} = 4.4$, and $\kappa_{RD} = 5.0$, $\mu_{RD} = 1.2$, $m_{RD} = 2.8$  The simulation parameter are similar to \cite{Charishma2021OutageIRS}. We set $\alpha_{SD} = 2.0$, $\alpha_{SR} = 3.0$, and $\alpha_{RD} = 1.0$ unless mentioned otherwise. 
	In Fig. \ref{fig: OP_SISO_IRS_AKMS_VaryN_Exp1}, we plotted the OP for different values of $N$. It is evident from Fig. \ref{fig: OP_SISO_IRS_AKMS_VaryN_Exp1} that the approximation derived in Section \ref{SubSec: ProductAKMS_SISO_IRS} matches with simulated values. Also, as the number of IRS elements increases, the OP decreases. Next, in Fig. \ref{fig: OP_SISO_IRS_AKMS_Vary_aSD_Exp1}, we plotted the OP for different $\alpha_{SD}$. Interestingly, the higher value of $\alpha_{SD}$ has a positive impact only up to a certain threshold. After that, we observed the opposite behavior. We can see in the figure that for $\gamma_{th} < 0$ dB, OP is less for higher $\alpha_{SD}$ value, but for $\gamma_{th} > 0$ dB; OP is higher for higher $\alpha_{SD}$ values. 
	
	\begin{figure}
		\centering
		\includegraphics[width=0.45\textwidth]{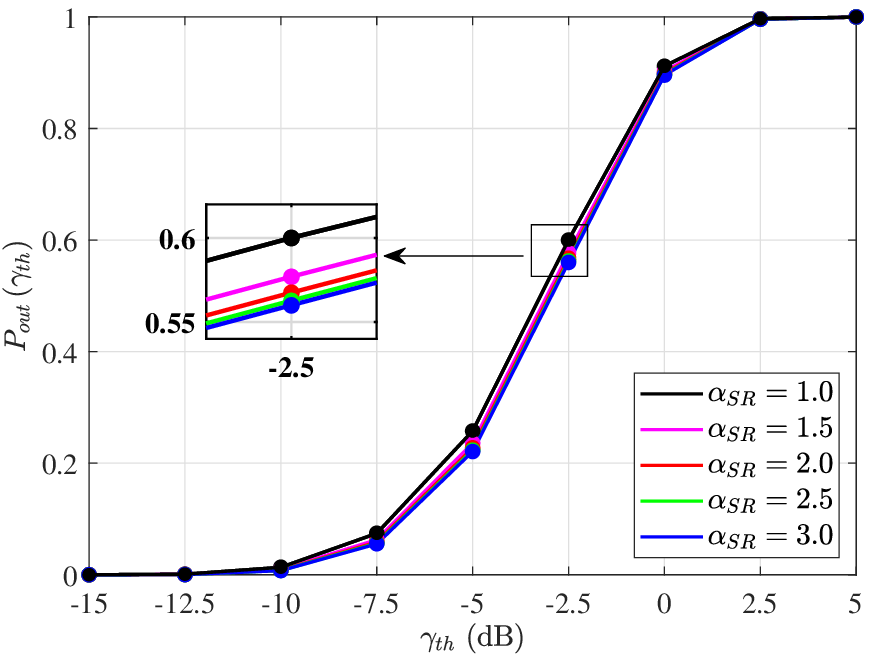}
		\captionof{figure}{OP of IRS-Assisted Communication System for various $\alpha_{SR}$}
		\label{fig: OP_SISO_IRS_AKMS_Vary_aSR_Exp1}
	\end{figure}
	
	\begin{figure}
		\centering
		\includegraphics[width=0.45\textwidth]{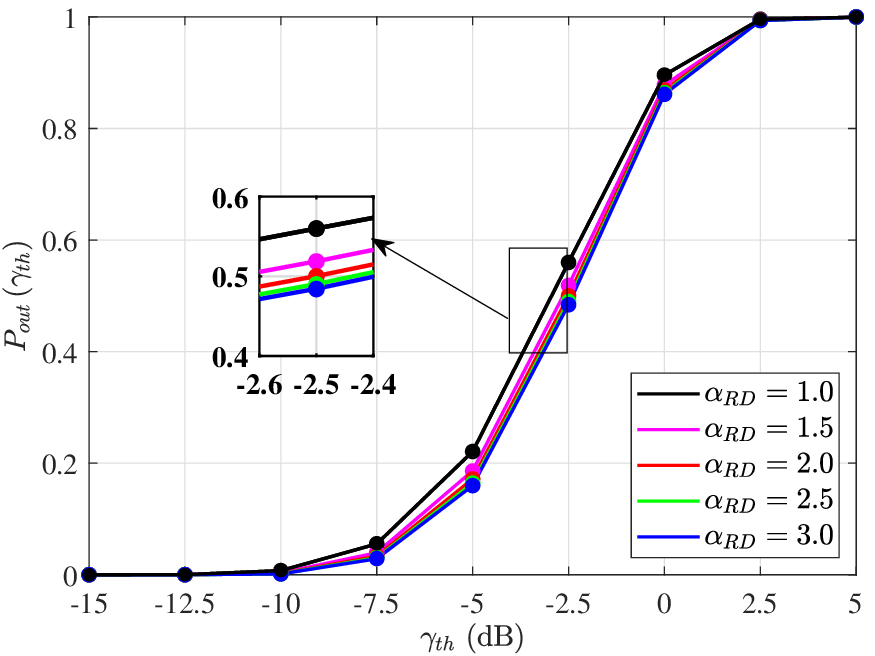}
		\captionof{figure}{OP of IRS-Assisted Communication System for various $\alpha_{RD}$}
		\label{fig: OP_SISO_IRS_AKMS_Vary_aRD_Exp1}
	\end{figure}

	Fig. \ref{fig: OP_SISO_IRS_AKMS_Vary_aSR_Exp1} and \ref{fig: OP_SISO_IRS_AKMS_Vary_aRD_Exp1} presents the OP for different values of $\alpha_{SR}$ and $\alpha_{RD}$, respectively. We can observe that the OP decreases as $\alpha_{SR}$ or $\alpha_{RD}$ increases. This is justified as we have seen in Fig. \ref{fig: AF_Cascade_AKMS_Vary_a} that the amount of fading decreases for a cascade channel as the $\alpha$ parameter increases for either of the links.

	\section{Conclusion}\label{Sec: ProductAKMS_Conclusion}
	This paper presents the series expression for PDF, CDF, and MGF of the product and ratio of two $ \alpha-\kappa-\mu $ shadowed RV. The series expression is obtained via a direct application of Mellin transformation. We have also derived a Laplace-type integral representation for the product and ratio statistics PDF. Some typical examples of wireless communication applications, like cascaded systems and physical layer security, are provided. While we have provided a 6G application, namely an IRS-assisted communication system, the fading model considered in this paper is quite general and encompasses typically used fading models. Hence, we hope our results can be used for other emerging applications. These theoretical results are important not only to assess the performance of the considered application but also helpful in proper resource allocation. An interesting extension for this work is considering the product of an arbitrary number of $ \alpha-\kappa-\mu $ shadowed RVs.
	
	\bibliographystyle{IEEEtran}
	\bibliography{RefAlphaKMS}

\end{document}